\documentclass[a4paper,12pt]{article}

\usepackage{fullpage}
\usepackage{amssymb}

\usepackage{mathtools}
\usepackage{rotating}
\usepackage[latin1]{inputenc}
\usepackage{graphicx}
\usepackage{natbib}
\usepackage{url}
\usepackage{amsmath}
\usepackage{amsfonts}
\usepackage{url}
\usepackage{color}
\usepackage{subfigure}
\usepackage{slashbox}
\usepackage{pstricks,pst-node,pst-tree}
\usepackage{ctable}

\def\bmu{\boldsymbol{\mu}}

\begin{document}

\title{Hypothesis Testing for Parsimonious\\ Gaussian Mixture Models} 

\author{
Antonio Punzo
\thanks{
	Department of Economics and Business,
	University of Catania,
	Corso Italia 55, 95129 Catania, Italy.
	Tel.: +39-095-7537640,
	e.mail: \texttt{antonio.punzo@unict.it}
}
\qquad
Ryan P.\ Browne
\thanks{
	Department of Mathematics \& Statistics, 
	University of Guelph,
	Guelph, Ontario, Canada, N1G 2W1.
  Tel.: +1-519-8244120, ext.\ 53034, 
  e.mail: \texttt{rbrowne@uoguelph.ca}
	}
\qquad
Paul D.\ McNicholas
\thanks{
	Department of Mathematics \& Statistics, 
	University of Guelph,
	Guelph, Ontario, Canada, N1G 2W1.
  Tel.: +1-519-8244120, ext.\ 53136, 
  e.mail: \texttt{paul.mcnicholas@uoguelph.ca}
	}	
}

\date{}

\maketitle

\begin{abstract}

Gaussian mixture models with eigen-decomposed covariance structures make up the most popular family of mixture models for clustering and classification, i.e., the Gaussian parsimonious clustering models (GPCM). Although the GPCM family has been used for almost 20 years, selecting the best member of the family in a given situation remains a troublesome problem. Likelihood ratio tests are developed to tackle this problems. These likelihood ratio tests use the heteroscedastic model under the alternative hypothesis but provide much more flexibility and real-world applicability than previous approaches that compare the homoscedastic Gaussian mixture versus the heteroscedastic one. 
Along the way, a novel maximum likelihood estimation procedure is developed for two members of the GPCM family.
Simulations show that the $\chi^2$ reference distribution gives reasonable approximation for the LR statistics only when the sample size is considerable and when the mixture components are well separated; accordingly, following \citet{Lo:Alik:2008}, a parametric bootstrap is adopted.
Furthermore, by generalizing the idea of \citet{Gres:Punz:Clos:2013} to the clustering context, a closed testing procedure, having the defined likelihood ratio tests as local tests, is introduced to assess a unique model in the general family.
The advantages of this likelihood ratio testing procedure are illustrated via an application to the well-known Iris data set.

\end{abstract}


\section{Introduction}
\label{sec:Introduction}

The Gaussian mixture model (see Sect.~\ref{sec:Mixtures of Gaussian distributions}) has been extensively considered as a powerful device for clustering by typically assuming that each mixture component represents a group (or cluster or class) in the original data (cf.\ \citealp{Titt:Smit:Mako:stat:1985}, \citealt{Fral:Raft:Howm:1998}, and \citealp{McLa:Basf:mixt:1988}); however, merging can also be considered to allow more than one component to represent a class \citep[e.g.,][]{hennig10}. Its popularity is largely attributable to its computational and theoretical convenience, as well as the speed with which it can be implemented for many data sets.
Attention on Gaussian mixtures significantly increased since the work of \citet{Cele:Gova:Gaus:1995}, who proposed a family of fourteen Gaussian parsimonious clustering models (GPCMs) obtained by imposing some constraints on eigen-decomposed component covariance matrices. Popular software soon emerged for efficient implementation of some members of the GPCM family and severed to further bolster their popularity \citep[cf.][]{Fral:Raft:Mode:2002}.

The GPCM family can be regarded as containing three subfamilies: the \textit{spherical family} with two members that have spherical components, the \textit{diagonal family} composed by four members that have axis-aligned components, and the \textit{general family} with eight members that generate more flexible components.
Homoscedasticity and heteroscedasticity represent the extreme configurations, in parsimonious terms, in the general family (see Sect.~\ref{sec:Mixtures of Gaussian distributions}).
\citet{Cele:Gova:Gaus:1995} describe maximum likelihood (ML) estimation for the models in the general family (see Sect.~\ref{subsec:ML}); however, for two of these models, the authors relax one of the assumptions on which the family is based on, i.e., the assumption of decreasing order of the eigenvalues on the diagonal of the eigenvalues matrix.
To overcome this problem, ML parameter estimation under order constraints is here proposed and illustrated (see Sect.~\ref{subsec:ML for EVE and VVE}).
  
When the number of components is either known \textit{a priori} or determined by some of the methods available in the literature \citep[see][Chapt.~6 and the references therein]{McLa:Peel:fini:2000}, a likelihood-ratio (LR) statistic can be used for comparing the models in the general family.
Unfortunately, attention has focused solely on the comparison between homoscedastic and heteroscedastic Gaussian mixtures and is further restricted to the univariate case \citep{Lo:Alik:2008}.
Herein, LR tests adopting the heteroscedastic Gaussian mixture model under the alternative hypothesis are considered for all the members of the GPCM family (Sect.~\ref{sec:Testing in the general family}).
For these tests, simulation results show that the $\chi^2$ reference distribution gives reasonable approximation for the LR statistic only when the sample size is considerable and when the mixture components are well separated (Sect.~\ref{subsec:Null distribution of the LR statistic}). This is expected within the mixture modelling context.
In line with \citet{Lo:Alik:2008}, a parametric bootstrap approach is so presented to approximate the distribution of the LR statistic (Sect.~\ref{subsec:Parametric bootstrap LR tests}).

One drawback with the tests discussed above is that they are only pairwise tests, i.e., each model in the general family is separately compared with the benchmark heteroscedastic Gaussian mixture. An ``overall'' testing procedure that detects the model by simultaneously considering all the members of the general family is preferable.
With this in mind, a closed testing procedure is developed based on the defined LR tests and building on recent work by  \citet{Gres:Punz:Clos:2013} (Sect.~\ref{sec:Testing in the general family}).
Computational aspects related to the implementation of the single LR test, and also to the implementation of the closed testing procedure, are given in Sect.~\ref{sec:Computational aspects}.
The well-known Iris data set is considered in Sect.~\ref{sec:Analysis on the Iris data} to illustrate the procedure and to demonstrate its advantages. 
 
\section[Parsimonious Gaussian mixtures]{The GPCM Family}
\label{sec:Parsimonious mixtures of Gaussian distributions}

\subsection[Mixtures of Gaussian distributions]{The family}
\label{sec:Mixtures of Gaussian distributions}

The distribution of a $p$-variate random vector $\boldsymbol{X}$ from a mixture of $k$ Gaussian distributions is
\begin{equation}
f\left(\boldsymbol{x};\boldsymbol{\vartheta}\right)=\sum_{j=1}^k\pi_j\phi\left(\boldsymbol{x};\boldsymbol{\mu}_j,\boldsymbol{\Sigma}_j\right),
\label{eq:Gaussian mixture}
\end{equation}
where $\pi_j>0$ is the mixing proportion of the $j$th component, with $\sum_{j=1}^k\pi_j=1$, $\phi\left(\boldsymbol{x};\boldsymbol{\mu}_j,\boldsymbol{\Sigma}_j\right)$ is the Gaussian density, with mean $\boldsymbol{\mu}_j$ and covariance matrix $\boldsymbol{\Sigma}_j$, and $\boldsymbol{\vartheta}=\left\{\pi_j,\boldsymbol{\mu}_j,\boldsymbol{\Sigma}_j\right\}_{j=1}^k$.

The Gaussian mixture model \eqref{eq:Gaussian mixture} can be overparametrized because there are $p\left(p + 1\right)/2$ free parameters for each $\boldsymbol{\Sigma}_j$. 
\citet{Banf:Raft:mode:1993} introduce parsimony  by considering the eigen-decomposition 
\begin{equation}
\boldsymbol{\Sigma}_j=\lambda_j\boldsymbol{\Gamma}_j\boldsymbol{\Delta}_j\boldsymbol{\Gamma}_j', 
\label{eq:eigenvalue decomposition}
\end{equation}
for $j=1,\ldots,k$, where $\lambda_j=\left|\boldsymbol{\Sigma}_j\right|^{1/p}$, $\boldsymbol{\Delta}_j$
is the scaled ($\left|\boldsymbol{\Delta}_j\right|=1$) diagonal matrix of the eigenvalues of $\boldsymbol{\Sigma}_j$ sorted in decreasing order, and $\boldsymbol{\Gamma}_j$ is a $p\times p$ orthogonal matrix whose columns are the normalized eigenvectors of $\boldsymbol{\Sigma}_j$, ordered according to their eigenvalues.
Each element in the right side of \eqref{eq:eigenvalue decomposition} has a different geometric interpretation: $\lambda_j$ determines the volume of the cluster, $\boldsymbol{\Delta}_j$ its shape, and $\boldsymbol{\Gamma}_j$ its orientation. \citet{Cele:Gova:Gaus:1995} impose constraints on the elements on the right-hand side of \eqref{eq:eigenvalue decomposition} to give a family of 14 Gaussian parsimonious clustering models. These 14 models include very specific special cases, e.g., $\boldsymbol{\Gamma}_j=\boldsymbol{I}$ (identity matrix), and more general constraints, e.g., $\boldsymbol{\Gamma}_j=\boldsymbol{\Gamma}$.

\begin{figure}[!ht] 
\centering
	\resizebox{0.95\textwidth}{!}{\includegraphics{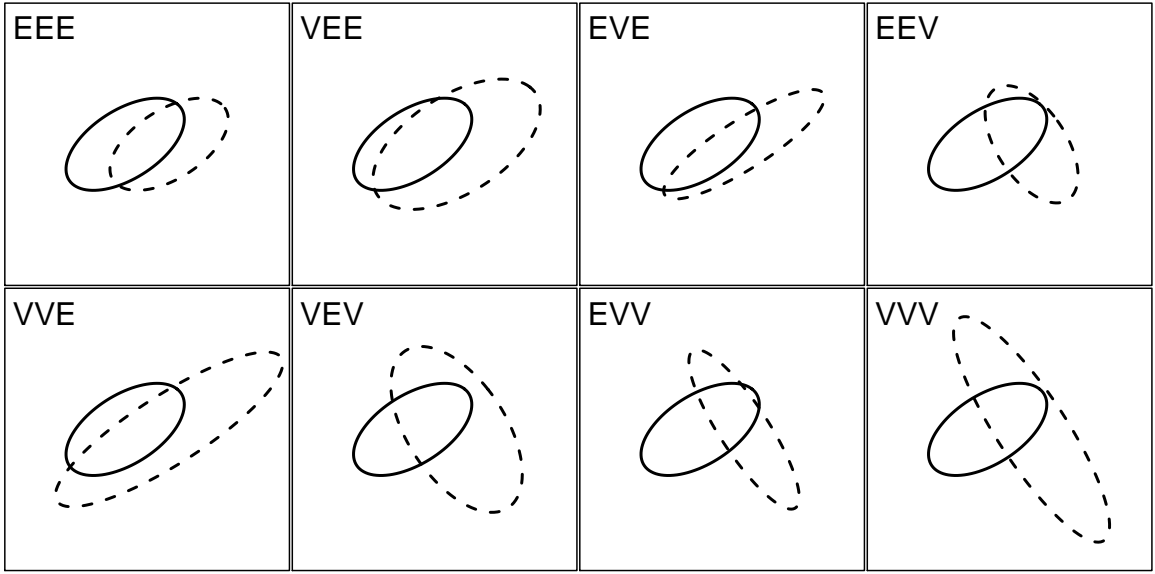}}
	\caption{
	Examples of the models in $\widetilde{\mathcal{M}}$ in the bivariate case ($p=2$) with $k=2$ components.
	}
	\label{fig:francobolli}
\end{figure}

Herein, we focus on the more general constraints. To this end, consider the triplet $\left(\lambda_j,\boldsymbol{\Delta}_j,\boldsymbol{\Gamma}_j\right)$ and allow its elements to be equal (E) or variable (V) across components. This leads to a `general family' $\widetilde{\mathcal{M}}$ of eight models detailed in \tablename~\ref{tab:general family}.
With this convention, writing EEV means that we consider groups with equal volume, equal shape, and different orientation. 
\begin{table}[!ht]
\caption{
\label{tab:general family}
Models in the general family $\widetilde{\mathcal{M}}$ described by their covariance restrictions.}
\centering
\resizebox*{1\textwidth}{!}{
\begin{tabular}{ccccccc}
\hline
\noalign{\smallskip}
$M$ & Volume & Shape & Orientation & $\boldsymbol{\Sigma}_j$ & ML & Free covariance parameters \\
\noalign{\smallskip}
\hline
\noalign{\smallskip}
EEE & Equal    & Equal     & Equal        & $\lambda\boldsymbol{\Gamma}\boldsymbol{\Delta}\boldsymbol{\Gamma}'$ & CF & $p\left(p+1\right)/2$\\
VEE & Variable & Equal     & Equal        & $\lambda_j\boldsymbol{\Gamma}\boldsymbol{\Delta}\boldsymbol{\Gamma}'$ & IP & $k+p-1+p\left(p-1\right)/2$ \\
EVE & Equal    & Variable  & Equal        & $\lambda\boldsymbol{\Gamma}_j\boldsymbol{\Delta}\boldsymbol{\Gamma}_j'$ & IP & $1+k\left(p-1\right)+p\left(p-1\right)/2$ \\
EEV & Equal    & Equal     & Variable     & $\lambda\boldsymbol{\Gamma}\boldsymbol{\Delta}_j\boldsymbol{\Gamma}'$ & CF & $p+kp\left(p-1\right)/2$ \\
VVE & Variable & Variable  & Equal        & $\lambda_j\boldsymbol{\Gamma}_j\boldsymbol{\Delta}\boldsymbol{\Gamma}_j'$ & IP & $kp+p\left(p-1\right)/2$ \\
VEV & Variable & Equal     & Variable   & $\lambda_j\boldsymbol{\Gamma}\boldsymbol{\Delta}_j\boldsymbol{\Gamma}'$ & IP & $k+p-1+kp\left(p-1\right)/2$ \\
EVV & Equal    & Variable  & Variable   & $\lambda\boldsymbol{\Gamma}_j\boldsymbol{\Delta}_j\boldsymbol{\Gamma}_j'$ & CF & $1+k\left(p-1\right)+kp\left(p-1\right)/2$ \\
VVV & Variable & Variable  & Variable   & $\lambda_j\boldsymbol{\Gamma}_j\boldsymbol{\Delta}_j\boldsymbol{\Gamma}_j'$ & CF & $kp\left(p+1\right)/2$ \\
\noalign{\smallskip}
\hline
\end{tabular}
}
\end{table}
\figurename~\ref{fig:francobolli} exemplifies the models providing a graphical representation in the case $p=k=2$.
For each model $M\in\widetilde{\mathcal{M}}$, the parameters in \eqref{eq:Gaussian mixture} can be denoted by $\boldsymbol{\vartheta}_M=\left\{\pi_j,\bmu_j,\lambda_j^M,\boldsymbol{\Delta}_j^M,\boldsymbol{\Gamma}_j^M\right\}_{j=1}^k$.

\subsection{Maximum likelihood parameter estimation}
\label{subsec:ML}

\begin{figure}[!ht]
\centering
\resizebox{0.55\textwidth}{!}
{
\includegraphics{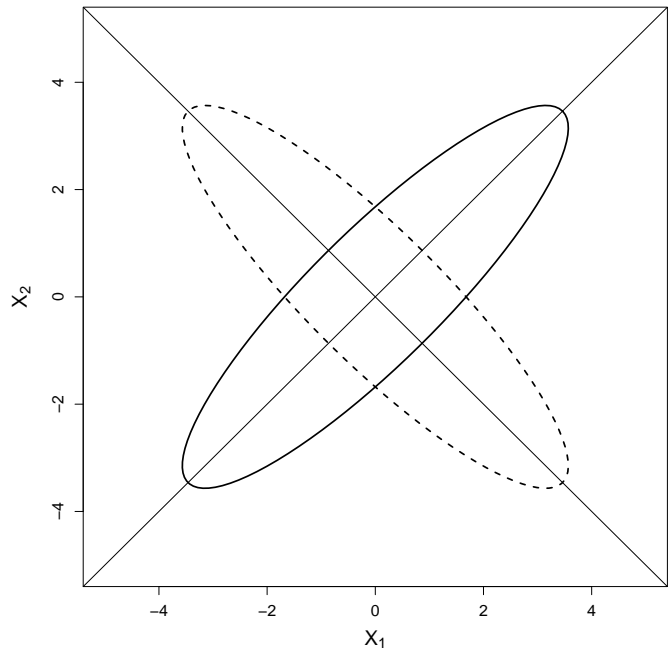}
}
\caption{ 
Ellipses related to the matrices in \tablename~\ref{tab:star}, where
components have mean $(0,0)'$.
\label{fig:EEVvsEEE}
}
\end{figure}

Given a sample $\boldsymbol{x}_1,\ldots,\boldsymbol{x}_n$ from model~\eqref{eq:Gaussian mixture}, once $k$ is assigned, the (observed-data) log-likelihood for the generic model $M\in\widetilde{\mathcal{M}}$ can be written as
\begin{equation}
l_M\left(\boldsymbol{\vartheta}_M\right)=\sum_{i=1}^n \log f\left(\boldsymbol{x}_i;\boldsymbol{\vartheta}_M\right).
\label{eq:observed-data log-likelihood}
\end{equation}
The EM algorithm of \citet{Demp:Lair:Rubi:Maxi:1977} can be used to maximize $l_M$ in order to find maximum likelihood (ML) estimates for $\boldsymbol{\vartheta}_M$.
The algorithm basically works on the complete-data log-likelihood, i.e., 
\begin{equation}
l_M^c\left(\boldsymbol{\vartheta}_M\right)=\sum_{i=1}^n \sum_{j=1}^k z_{ij}\left[\log \pi_j + \log \phi\left(\boldsymbol{x}_i;\boldsymbol{\mu}_j,\boldsymbol{\Sigma}_j^M\right)\right], 
\label{eq:complete-data log-likelihood}
\end{equation}
where $z_{ij}=1$ if $\boldsymbol{x}_i$ comes from component $j$, and $z_{ij}=0$ otherwise.
The penultimate column of \tablename~\ref{tab:number of parameters} indicates whether maximization, in the context of the generic M-step of the EM algorithm, can be achieved in a closed form (CF) or if an iterative procedure (IP) is needed; details can be found in \citet{Cele:Gova:Gaus:1995} and \citet[pp.~22--24]{Bier:Cele:Gova:Lang:Noul:Vern:MIXM:2008}.
However, for the models EVE and VVE, characterized by a common eigenvector matrix $\boldsymbol{\Gamma}$, \citet{Cele:Gova:Gaus:1995} only describe an M-step for the weaker assumption of ``equality in the set of $p$ eigenvectors between groups'' while, as stated in Sect.~\ref{sec:Mixtures of Gaussian distributions}, we need equality in the ordered set of $p$ eigenvectors between groups.
This is also a fundamental requirement for the general family to be closed (see Sect.~\ref{sec:Testing in the general family} for details).
To motivate this problem, we consider the EEV model in \tablename~\ref{tab:star} (see also \figurename~\ref{fig:EEVvsEEE}).
\begin{table}[!ht]
\caption{
Example of EEV model in the case of two groups in two dimensions.
\label{tab:star}
}
\centering
\resizebox{\textwidth}{!}{
\begin{tabular}{c|cccccc|c}
\toprule
& Volume && Shape && Orientation && Comment\\
\midrule
Group 1 & $\lambda_1=1$ && $\boldsymbol{\Delta}_1=
\begin{bmatrix}
4 & 0 \\ 
0 & 1/4 \\ 
\end{bmatrix}
$
&&
$\boldsymbol{\Gamma}_1=
\begin{bmatrix*}[r]
\sqrt{2}/2 & -\sqrt{2}/2 \\
\sqrt{2}/2 & \sqrt{2}/2 \\
\end{bmatrix*}
$
&&
\\[8mm]
Group 2 & $\lambda_2=1$ && 
$\boldsymbol{\Delta}_2=
\begin{bmatrix}
4 & 0 \\
0 & 1/4 \\ 
\end{bmatrix}
$
&&
$\boldsymbol{\Gamma}_2=
\begin{bmatrix*}[r]
-\sqrt{2}/2 & \sqrt{2}/2 \\
\sqrt{2}/2 & \sqrt{2}/2 \\
\end{bmatrix*}
$
&& 
\begin{tabular}{l}
Same ordering and values for the scaled\\ 
eigenvalues in $\boldsymbol{\Delta}_1$ and $\boldsymbol{\Delta}_2$ but $\boldsymbol{\Gamma}_1\neq\boldsymbol{\Gamma}_2$
\end{tabular}
\\
\bottomrule
\end{tabular}
}
\end{table}
This configuration erroneously corresponds to the EEE model in the ``modified'' general family of \citet{Cele:Gova:Gaus:1995}.
For a further discussion about this issue see \cite{Gres:Ingr:Punz:Asse:2011}, \citet{Gres:Punz:Clos:2013}, and \citet{Bagn:Gres:Punz:Onth:2014}.

\subsection{Maximum likelihood parameter estimation under order constraints for EVE and VVE models}
\label{subsec:ML for EVE and VVE}

Motivated by the example given at the end of Sect.~\ref{subsec:ML}, we extend estimation procedures  given in \citet{Cele:Gova:Gaus:1995} for the models EVE and VVE to the case where we require the between-group eigenvalues to have the same decreasing order. 

\subsubsection{VVE Model: $\boldsymbol{\Sigma}_j=\lambda_j \boldsymbol{\Gamma}\boldsymbol{\Delta}_j\boldsymbol{\Gamma}'$}
\label{sec:VVE Model}

If we let $\boldsymbol{\Xi}_j = \lambda_j\boldsymbol{\Delta}_j$, where $\boldsymbol{\Xi}_j=\mbox{diag}\left(\xi_{j1},\ldots, \xi_{jp}\right)$, then maximizing the complete-data log-likelihood \eqref{eq:complete-data log-likelihood} is equivalent to minimizing the $k$ optimization problems (one for each group $j$):
\begin{equation*}
\begin{aligned}
& \underset{\xi_{j1}, \ldots , \xi_{jp}}{\text{minimize}}
& &  \mbox{tr}\left( \boldsymbol{W}_j \boldsymbol{\Gamma}\boldsymbol{\Xi}_j^{-1}\boldsymbol{\Gamma}' \right) + n_j \log \left|  \boldsymbol{\Xi}_j \right|  \\
& \text{subject to}
& & \xi_{j1} \geq \xi_{j2} \geq \cdots \geq \xi_{jp} \ge 0, 
 \end{aligned}
\end{equation*}
where, for the $j$th group, $\boldsymbol{W}_j$ is the weighted scatter matrix and $n_j=\sum_{i=1}^nz_{ij}$, $j=1,\ldots,k$. 
This optimization problem is not convex because the second derivative of the objective function can be negative. 
However, if we apply the transform $\zeta_{jl} = \log(\xi_{jl})$, which is one-to-one and ordering preserving, we obtain the convex programming problem:
\begin{equation} 
\label{transformed convex program}
\begin{aligned}
& \underset{\zeta_{jl},\ldots, \zeta_{jp}}{\text{minimize}}
& & \sum_{l=1}^p \left[ b_{jl}  \exp\left(-\zeta_{jl}\right) + n_j \zeta_{jl} \right] \\
& \text{subject to}
& & \zeta_{j1} \geq \zeta_{j2} \geq \cdots \geq \zeta_{jp} , 
 \end{aligned}
\end{equation}
where $b_{jl} =  \boldsymbol{\gamma}_l'\boldsymbol{W}_j \boldsymbol{\gamma}_l$, with $\boldsymbol{\gamma}_l$ being the $l$th column vector of $\boldsymbol{\Gamma}$, also called the $l$th eigenvector.
Advantageously, this convex programming problem has linear constraints and if a set of constraints are known to be active then the solution is easily to obtain. 
Thus, the primal active set method \citep{nocedal2000} is a good algorithm to apply this problem. 
Then to update the common orientation matrix, $\boldsymbol{\Gamma}$, one can apply the methodology from \citet{Flur:Gaut:anal:1986} and \citet{Brow:McNi:Esti:2013,Brow:McNi:Orth:2012}. 

\subsubsection{EVE Model: $\boldsymbol{\Sigma}_j=\lambda \boldsymbol{\Gamma}\boldsymbol{\Delta}_j\boldsymbol{\Gamma}'$}
\label{sec:EVE Model}

For the case where the volume is equal across groups, we apply the same methodology from Sect.~\ref{sec:VVE Model}. However, in the transformed convex programming problem (\ref{transformed convex program}), for each $j$ we have an additional constraint that $\sum_{l=1}^p \zeta_{jl} = 0 $, which is equivalent to $\prod_{l=1}^p \xi_{jl}=1$. 
This additional constraint is linear and can be adapted in the primal active set method. 

\section[Likelihood-ratio tests]{Likelihood-ratio tests}
\label{sec:likelihood-ratio tests}

Let $\overline{\mathcal{M}}=\widetilde{\mathcal{M}}/\left\{\text{VVV}\right\}$.
For each $M\in \overline{\mathcal{M}}$, a natural way to test
\begin{displaymath}
H_0^M:\text{``data arise from $M$''}\quad versus\quad H_1^{\text{VVV}}:\text{``data arise from VVV''},
\end{displaymath}
consists of using the (generalized) likelihood-ratio (LR) statistic
\begin{equation}
\text{LR}_{M} = 
-2\left[
l_M\left(\widehat{\boldsymbol{\vartheta}}_M\right)
-
l_{\text{VVV}}\left(\widehat{\boldsymbol{\vartheta}}_{\text{VVV}}\right)\right],
\label{eq:LR statistic}
\end{equation}
where $\widehat{\boldsymbol{\vartheta}}_M$ and $\widehat{\boldsymbol{\vartheta}}_{\text{VVV}}$ are the ML estimators of $\boldsymbol{\vartheta}_M$ and $\boldsymbol{\vartheta}_{\text{VVV}}$ under the null and alternative hypotheses, respectively.
Under some regularity conditions and under $H_0^M$, $\text{LR}_{M}$ is commonly assumed asymptotically distributed as $\chi^2$ with $\nu_M=\eta_{\text{VVV}}-\eta_{M}$ degrees of freedom, where $\eta_{\text{VVV}}$ and $\eta_{M}$ denote the number of (free) parameters for VVV and $M$, respectively.
The value of $\nu_M$ is the gain in parsimony that could be achieved.
\tablename~\ref{tab:number of parameters} specifies the number of parameters $\eta_{M}$, and the degrees of freedom $\nu_M$, for each $M\in \widetilde{\mathcal{M}}$.
\begin{table}[!ht]
\caption{
\label{tab:number of parameters}
Scheme of computation of $\nu_M$, for the asymptotic $\chi^2$-approximation of $\text{LR}_M$, starting from $\eta_{\text{VVV}}$ and $\eta_M$, $M\in \widetilde{\mathcal{M}}$.}
\centering
\resizebox*{1\textwidth}{!}{
\begin{tabular}{c c c c c cc}
\toprule
$M$  & 
$\eta_{\text{VVV}}$ 
&  &
$\eta_M$ 
& &
$\nu_M$
\\
\midrule
EEE & $\left(k-1\right)+kp+k\displaystyle\frac{p(p+1)}{2}$ & $-$ & $\left(k-1\right)+kp+\displaystyle\frac{p(p+1)}{2}$ & $=$ & $\left(k-1\right)\displaystyle\frac{p(p+1)}{2}$ \\[3mm]
VEE & $\left(k-1\right)+kp+k\displaystyle\frac{p(p+1)}{2}$ & $-$ & $\left(k-1\right)+kp+\displaystyle\frac{p(p+1)}{2}+\left(k-1\right)$ & $=$ & $\left(k-1\right)\left(\displaystyle\frac{p(p+1)}{2}-1\right)$ \\[3mm]
EVE & $\left(k-1\right)+kp+k\displaystyle\frac{p(p+1)}{2}$ & $-$ & $\left(k-1\right)+kp+\displaystyle\frac{p(p+1)}{2}+\left(k-1\right)\left(p-1\right)$ & $=$ & $\left(k-1\right)\left(\displaystyle\frac{p(p-1)}{2}+1\right)$\\[3mm]
EEV & $\left(k-1\right)+kp+k\displaystyle\frac{p(p+1)}{2}$ & $-$ & $\left(k-1\right)+kp+k\displaystyle\frac{p(p+1)}{2}-\left(k-1\right)p$ & $=$ & $\left(k-1\right)p$ \\[3mm]
VVE & $\left(k-1\right)+kp+k\displaystyle\frac{p(p+1)}{2}$ & $-$ & $\left(k-1\right)+kp+\displaystyle\frac{p(p+1)}{2}+\left(k-1\right)p$ & $=$ & $\left(k-1\right)\displaystyle\frac{p(p-1)}{2}$\\[3mm]
VEV & $\left(k-1\right)+kp+k\displaystyle\frac{p(p+1)}{2}$ & $-$ & $\left(k-1\right)+kp+k\displaystyle\frac{p(p+1)}{2}-\left(k-1\right)\left(p-1\right)$ & $=$ & $\left(k-1\right)\left(p-1\right)$\\[3mm]
EVV & $\left(k-1\right)+kp+k\displaystyle\frac{p(p+1)}{2}$ & $-$ & $\left(k-1\right)+kp+k\displaystyle\frac{p(p+1)}{2}-\left(k-1\right)$ & $=$ & $\left(k-1\right)$\\[3mm]
VVV & $\left(k-1\right)+kp+k\displaystyle\frac{p(p+1)}{2}  $ & $-$ & $\left(k-1\right)+kp+k\displaystyle\frac{p(p+1)}{2}$& $=$ &$0$\\
\bottomrule
\end{tabular}
}
\end{table}

\subsection[Null distribution of the LR statistic]{Null distribution of the LR statistic}
\label{subsec:Null distribution of the LR statistic}

Unfortunately, with mixture models, regularity conditions may not hold for $\text{LR}_{M}$, $M\in\overline{\mathcal{M}}$, to have the assumed $\chi^2$ reference distribution \citep{Lo:Alik:2008}.
Simulations are here conducted to examine this aspect.
Because many factors come into play (e.g., the number of groups $k$, the dimension $p$ of the observed variables, the overall sample size $n$, the volume, shape, and orientation elements of the eigen-decomposition), some of them are necessarily considered fixed for our purposes.

One thousand data sets are generated from each model in $\overline{\mathcal{M}}$.
We fix: $p=2$, $k=2$, $\pi_1=\pi_2=0.5$, and $\boldsymbol{\mu}_1=\boldsymbol{0}$.
With regard to the remaining parameters of the models, in the bivariate case, we have
\begin{displaymath}
\boldsymbol{\Sigma}_j=\lambda_j\boldsymbol{\Gamma}_j\boldsymbol{\Delta}_j\boldsymbol{\Gamma}_j'=
\lambda_j
\boldsymbol{R}\left(\gamma_j\right)
\begin{pmatrix}
1/\delta_j & & 0          \\[1ex]
           0 & & \delta_j
\end{pmatrix}
\boldsymbol{R}\left(\gamma_j\right)',
\end{displaymath}
where
\begin{displaymath}
\boldsymbol{R}\left(\gamma_j\right)=\begin{pmatrix}
\cos\gamma_j & &-\sin\gamma_j \\[1ex]
\sin\gamma_j & &\cos\gamma_j
\end{pmatrix}	
\end{displaymath}
is the rotation matrix of angle $\gamma_j$, and $\delta_j \in \left(0,1\right]$.
Note that the elements in the shape matrix arise from the constraint $\left|\boldsymbol{\Delta}_j\right|=1$.
Hence, we have a single parameter for each element of the eigen-decomposition: $\lambda_j$ is the volume parameter, $\delta_j$ is the shape parameter, and $\gamma_j$ is the orientation parameter (for further details see \citealp{Gres:Ingr:Punz:Asse:2011}, and \citealp{Gres:Punz:Clos:2013}). 
To generate data from each model, we preliminarily set $\boldsymbol{\Sigma}_1$ according to the values $\lambda_1=1$, $\delta_1=0.7$, and $\gamma_1=\pi/6$ (i.e., $30^{\circ}$).
With regard to $\boldsymbol{\Sigma}_2$, we choose $\lambda_2=3$ for models with variable volume, $\delta_2=0.3$ for models with variable shape, and $\gamma_2=\pi/6+\pi/4$ (i.e., $30^{\circ}+45^{\circ}$) for models with variable orientation.  
The second variate $\mu_{22}$ of $\boldsymbol{\mu}_2=\left(0,\mu_{22}\right)'$ is computed, via a numerical procedure, to guarantee a fixed overlap between groups.
Following \citet{Gres:Punz:Clos:2013}, we adopted the normalized measure of overlap 
\begin{equation*}
B=\exp\left(-B^*\right),
\label{eq:measure of overlap}
\end{equation*}
which takes values between 0 (absence of overlap) and 1 (complete overlap), where 
\begin{displaymath}
B^*=\frac{1}{8}\delta\left(\boldsymbol{\mu}_1,\boldsymbol{\mu}_2;\boldsymbol{\Sigma}^M\right)+\frac{1}{2}\log\left(\frac{\left|\boldsymbol{\Sigma}^M\right|}{\sqrt{\left|\boldsymbol{\Sigma}_1^M\right|+\left|\boldsymbol{\Sigma}_2^M\right|}}\right)	
\end{displaymath}
is the (positive) measure of overlap of \citet{Bhat:Onam:1943}, with $\boldsymbol{\Sigma}^M=\left(\boldsymbol{\Sigma}_1^M+\boldsymbol{\Sigma}_2^M\right)/2$.  
In particular, we consider three scenarios: $B=0.05$, $B=0.25$, and $B=0.45$. 
Three values for the sample size are also used: $n=100$, $n=200$, and $n=500$.
All nine combinations of the factors $B$ and $n$ are taken into account in the simulations.

\begin{figure}[!ht]
	\centering
		\resizebox{\textwidth}{!}{
		\begin{tabular}{c@{\hspace{-0.05mm}}c@{\hspace{-2mm}}c@{\hspace{-2mm}}c} 
		& $n=100$ & $n=200$ & $n=500$ \\[3mm]
		\begin{rotate}{90} \hspace{-0.4cm} $B=0.05$ \end{rotate}& 
		\begin{tabular}{c}
		\includegraphics[scale=0.33]{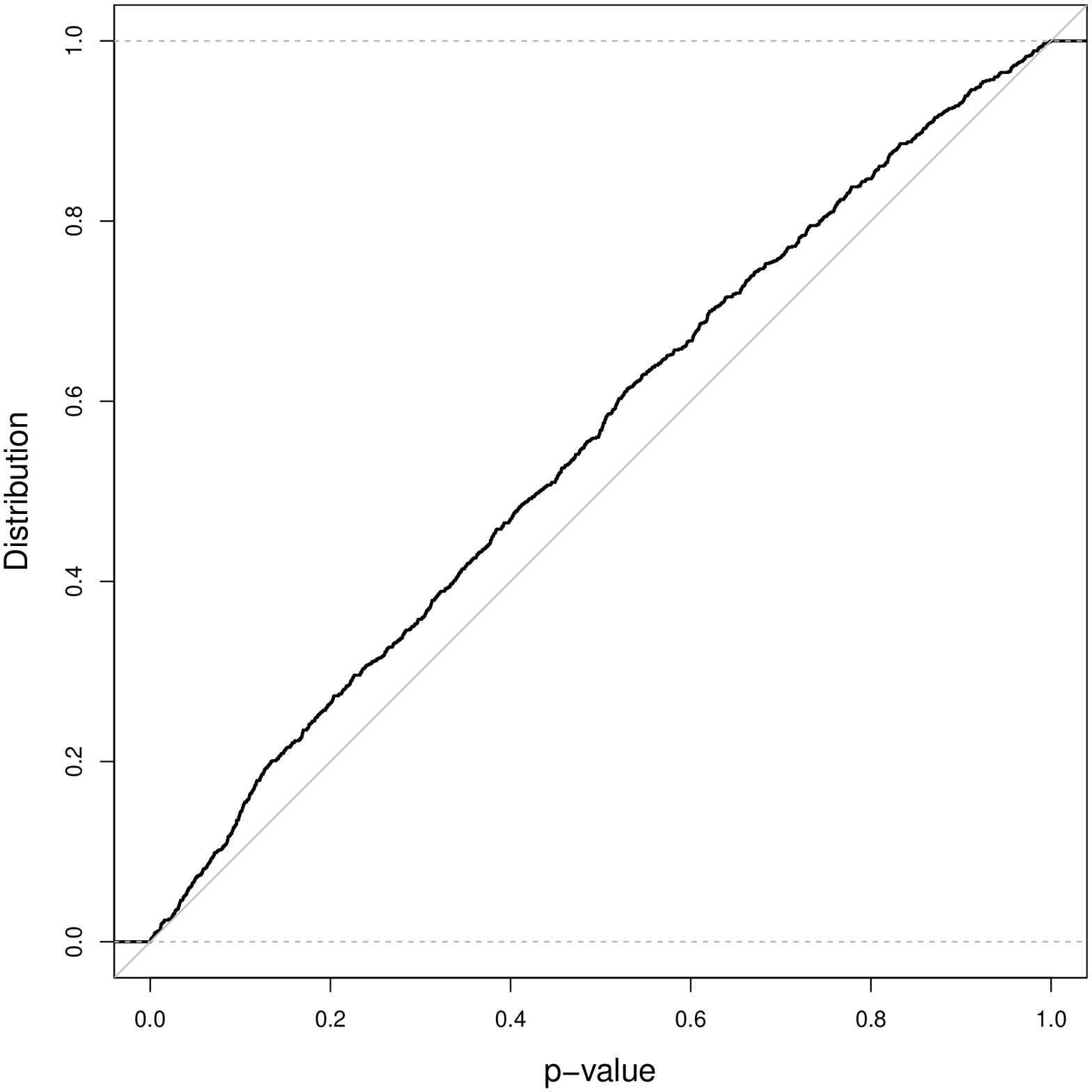} 
		\end{tabular}
		& 
		\begin{tabular}{c}
		\includegraphics[scale=0.33]{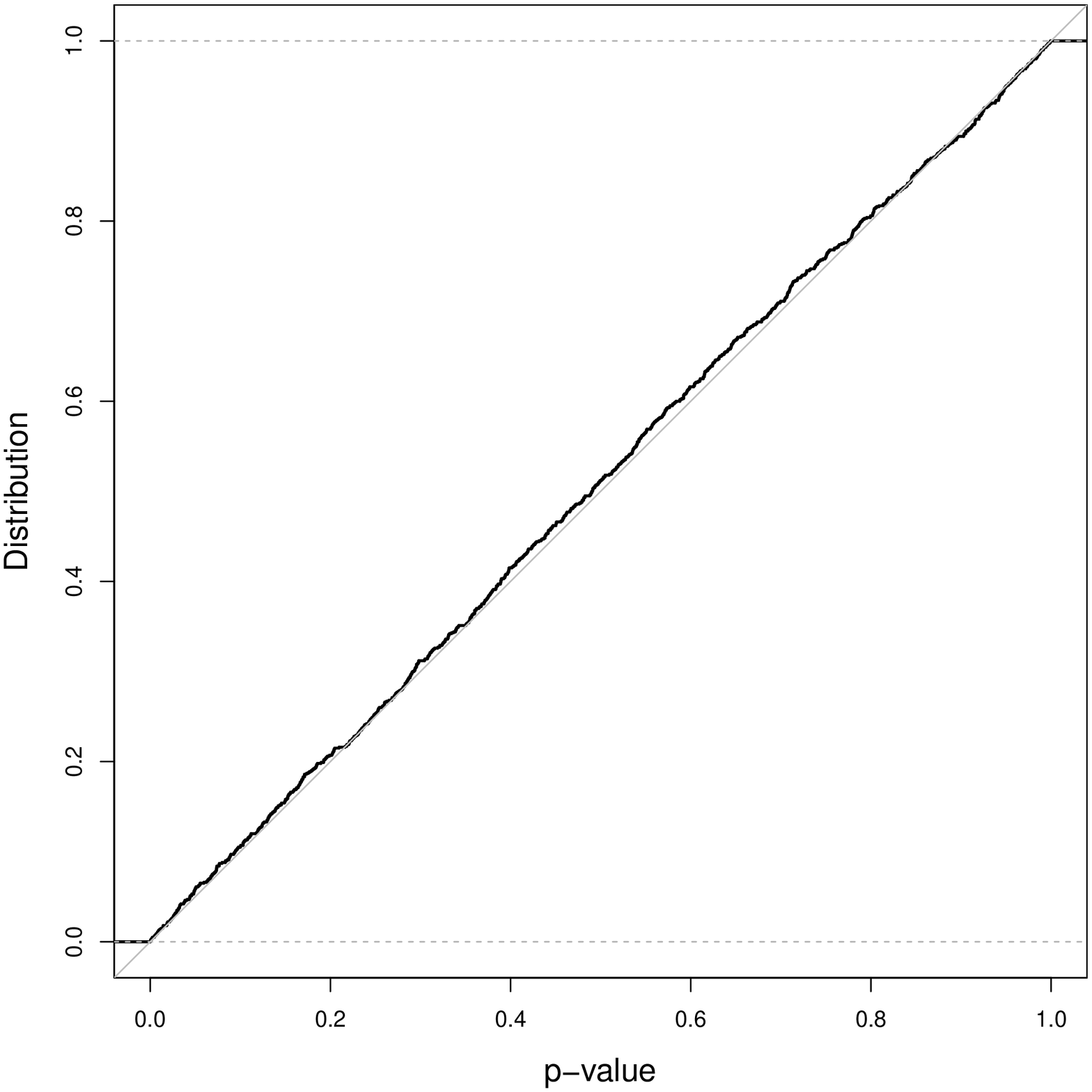} 
		\end{tabular}
		& 
		\begin{tabular}{c}
		\includegraphics[scale=0.33]{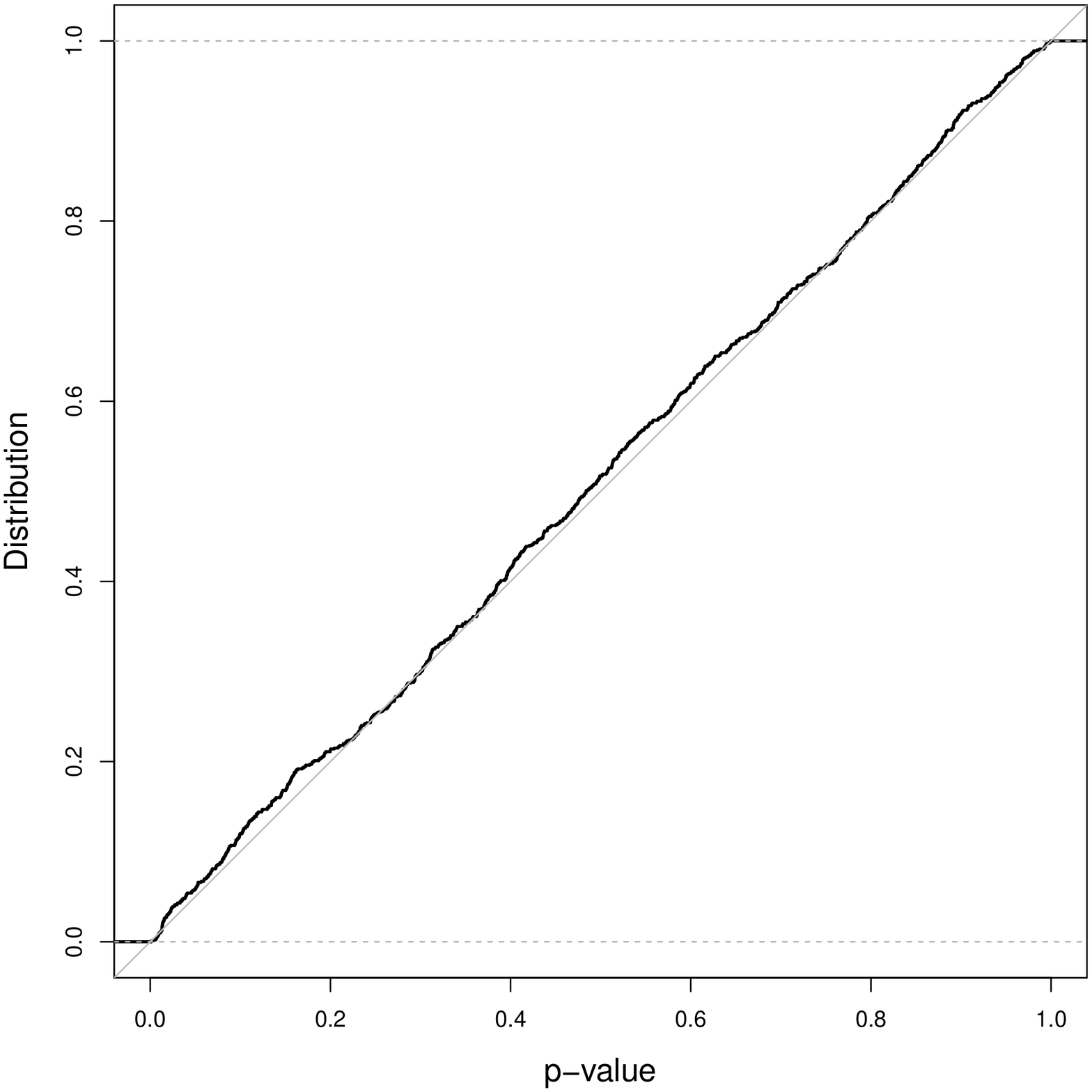} 
		\end{tabular}
		\\
		\begin{rotate}{90} \hspace{-0.39cm} $B=0.25$ \end{rotate}& 
		\begin{tabular}{c}
		\includegraphics[scale=0.33]{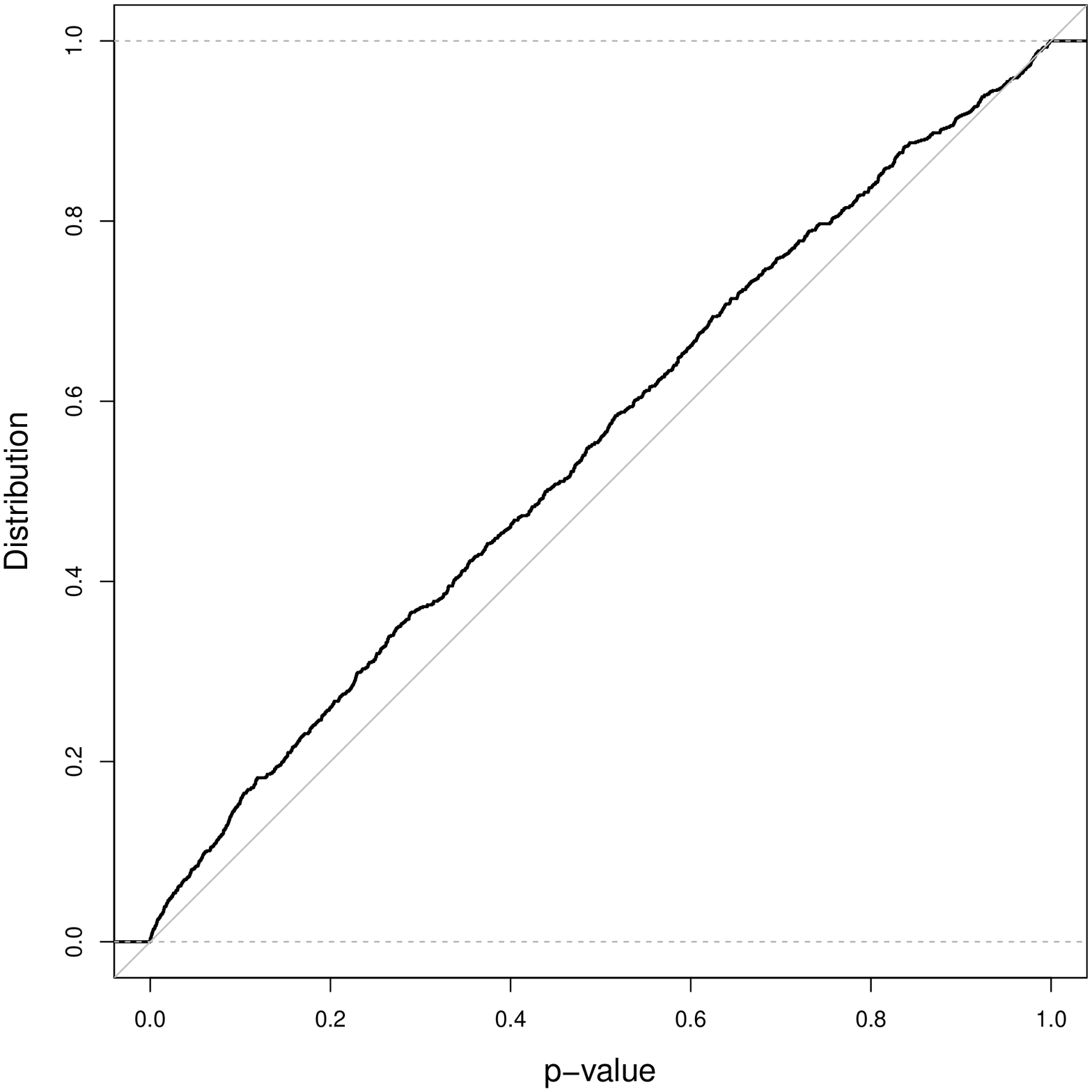} 
		\end{tabular}
		& 
		\begin{tabular}{c}
		\includegraphics[scale=0.33]{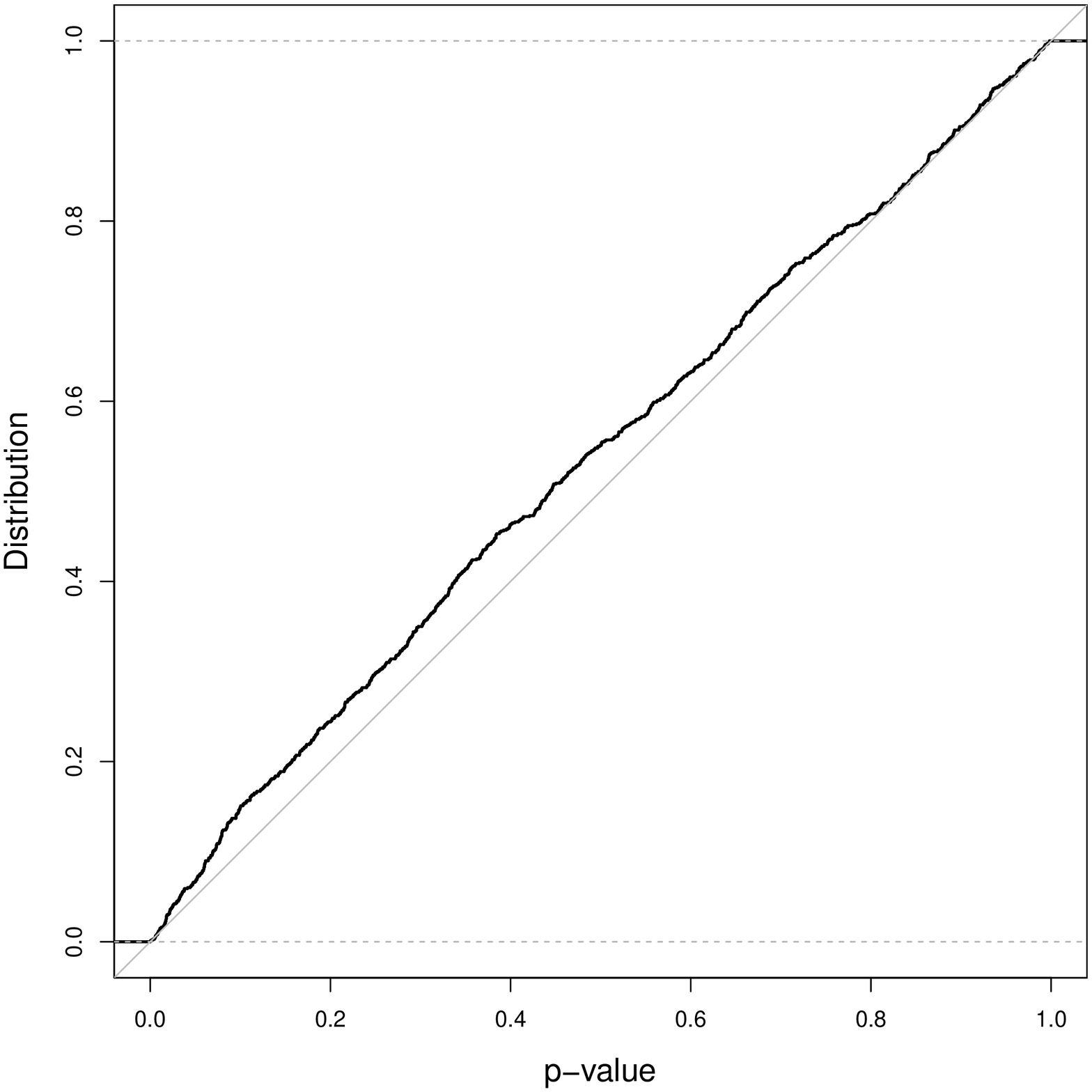} 
		\end{tabular}
		& 
		\begin{tabular}{c}
		\includegraphics[scale=0.33]{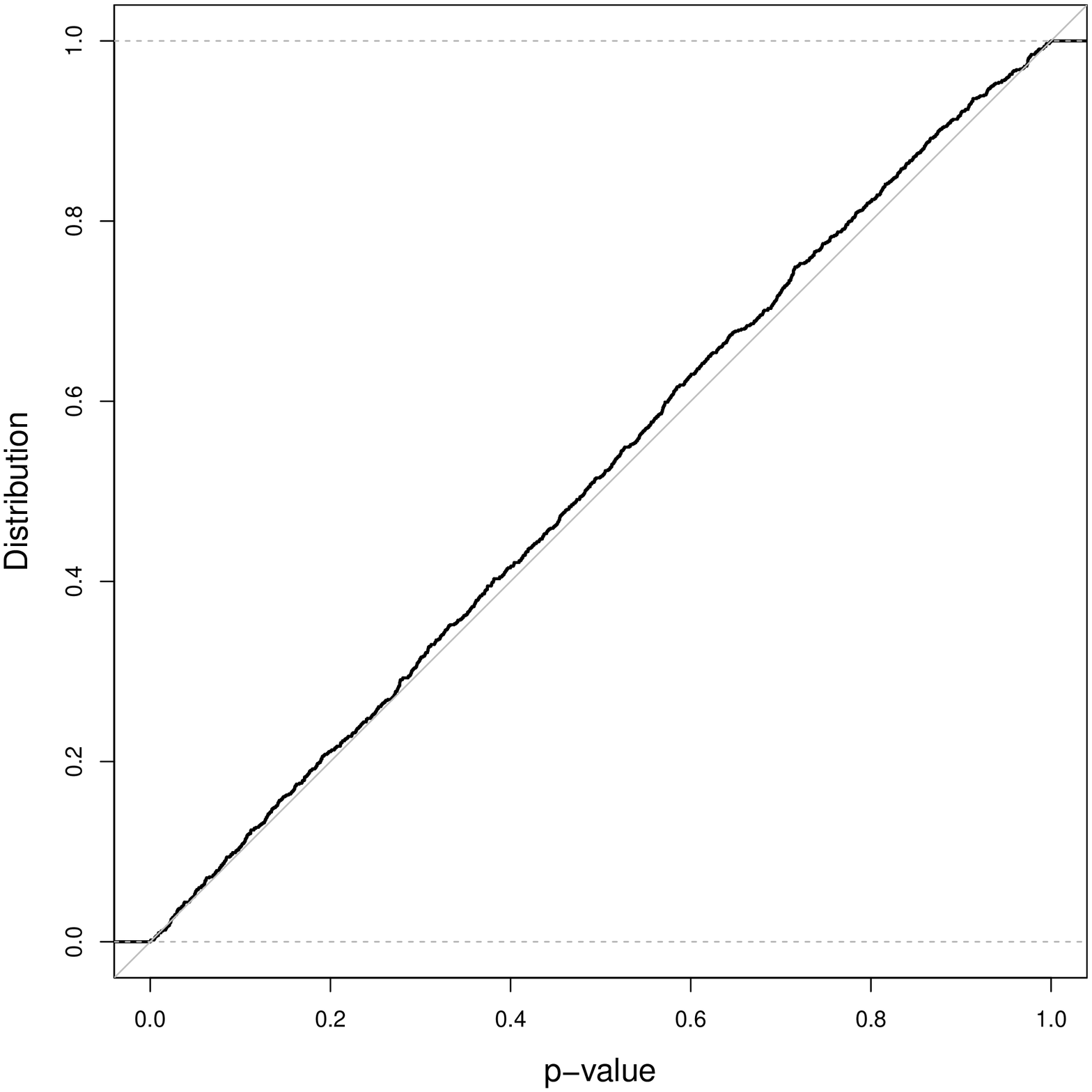} 
		\end{tabular}
		\\
		\begin{rotate}{90} \hspace{-0.40cm} $B=0.45$ \end{rotate}& 
		\begin{tabular}{c}
		\includegraphics[scale=0.33]{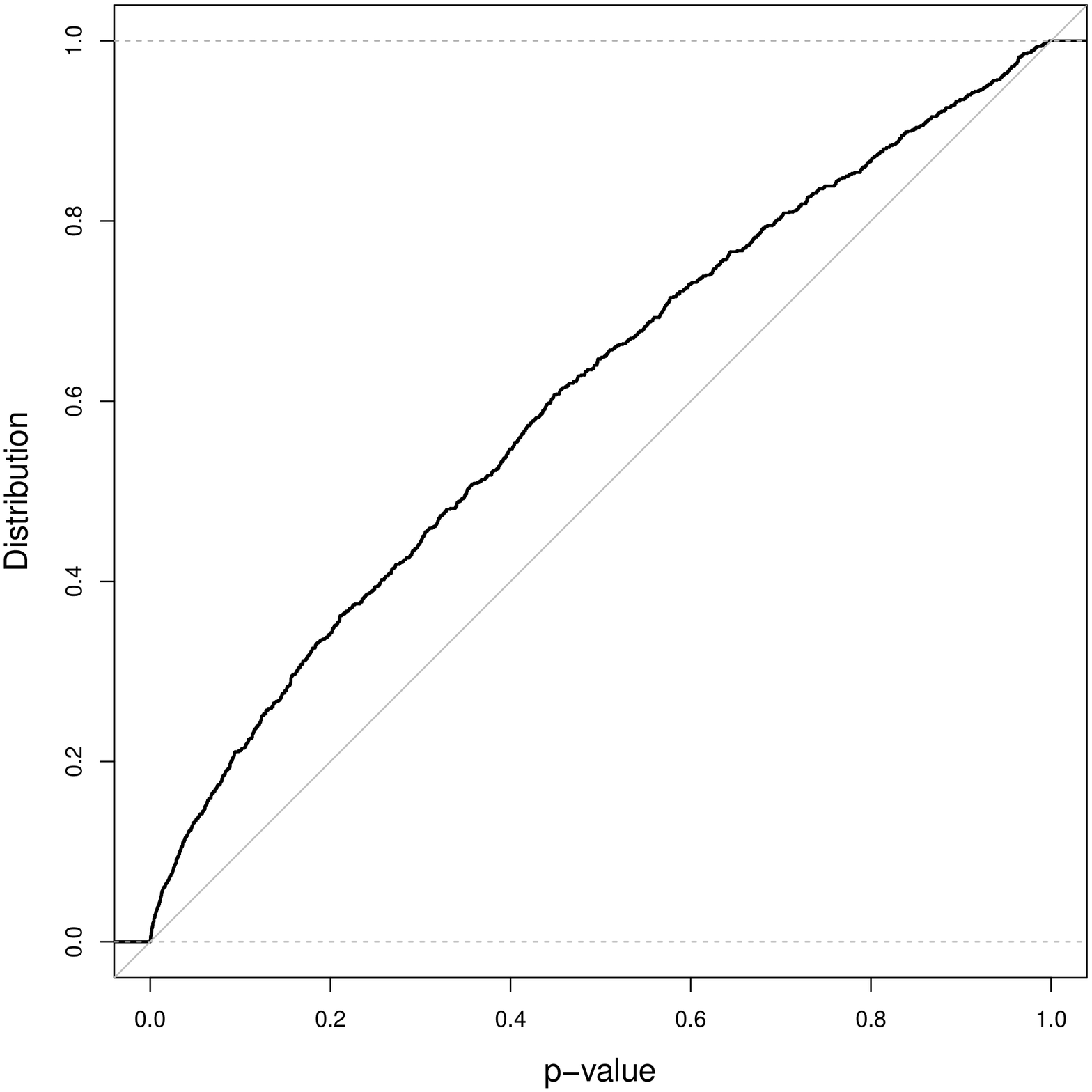} 
		\end{tabular}
		& 
		\begin{tabular}{c}
		\includegraphics[scale=0.33]{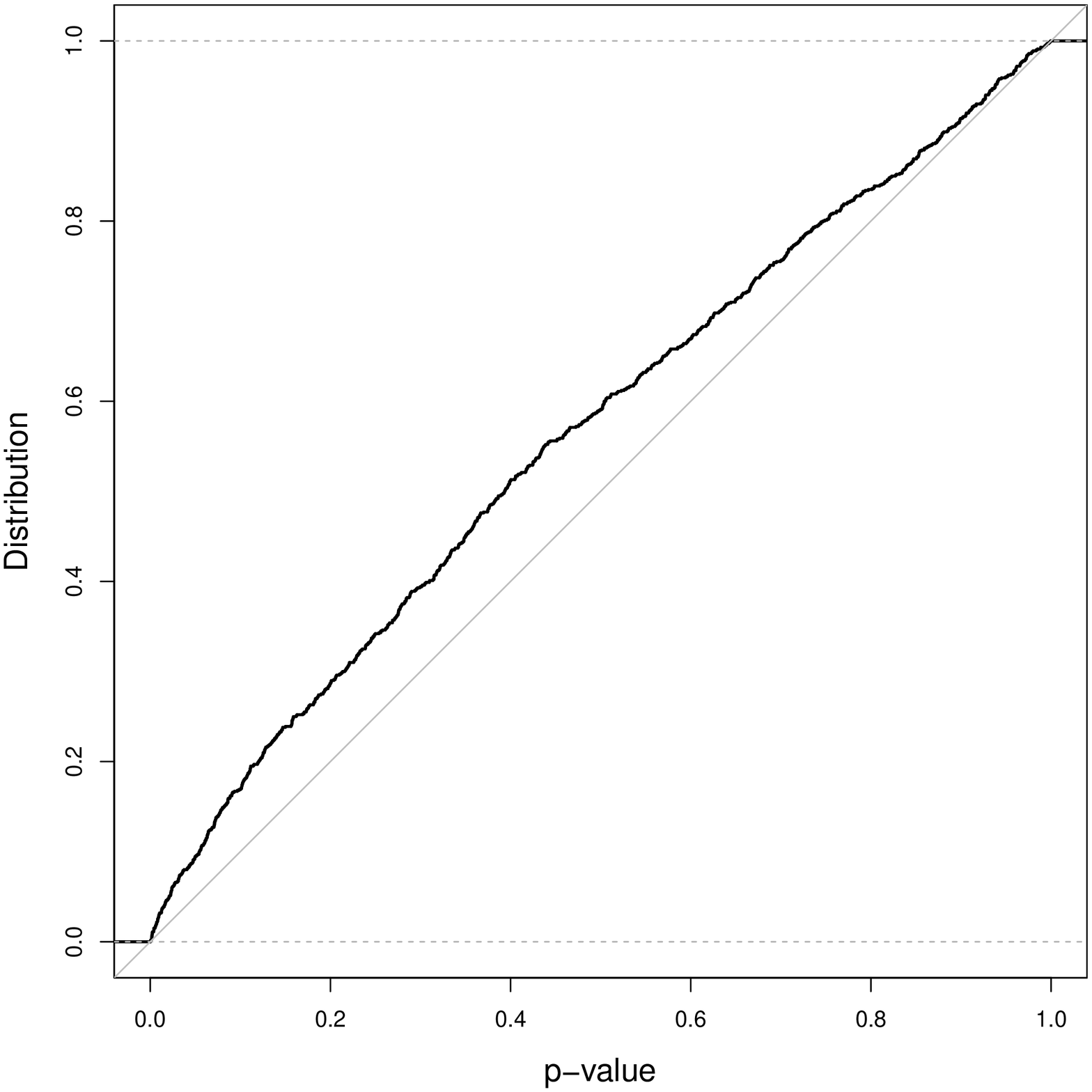} 
		\end{tabular}
		& 
		\begin{tabular}{c}
		\includegraphics[scale=0.33]{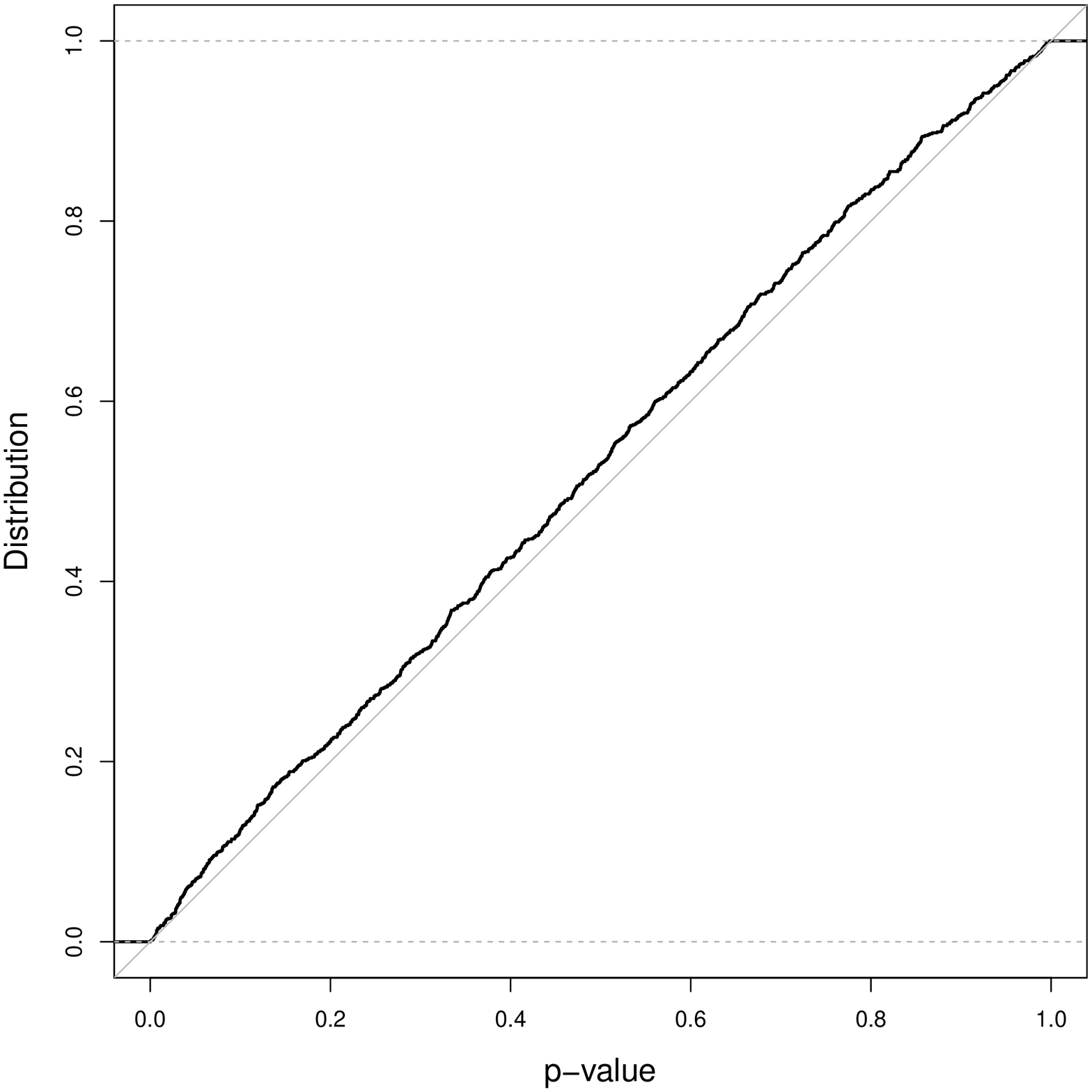} 
		\end{tabular}
	\end{tabular}
		}
	\caption{
	\footnotesize
	Asymptotic reference $\chi^2$ distribution: simulated distribution function of the $p$-values, for model EEE, at the varying of sample size $n$ and degree of overlap $B$.
	}
	\label{fig:EEE chi2}
\end{figure}

Because the results are similar across models, \figurename~\ref{fig:EEE chi2} shows the simulated distribution function (SDF) of the $p$-values (computed on 1000 replications) for EEE model only.
The obtained results are arranged as a matrix of plots where $n$ increases moving from left to right, while $B$ increases moving from top to bottom. 
In each subplot, if the null distribution is well approximated by the $\chi^2$ reference distribution, then we expect an SDF of the $p$-values very close to the distribution function of a uniform on $\left[0,1\right]$, superimposed in gray in each subplot of \figurename~\ref{fig:EEE chi2}. 
These results suggest that the assumed $\chi^2$ reference distribution gives a good approximation when the sample size increases and/or when the degree of overlap decreases.

\subsection{Parametric bootstrap LR tests}
\label{subsec:Parametric bootstrap LR tests}

As the null $\chi^2$ distribution does not provide a reasonable approximation for the LR statistics for small sample sizes (often encountered in practice) and for large overlap between groups, bootstrap methods that use the same criterion to compute $\text{LR}_M$ for each bootstrap re-sample can be used to approximate the sampling distribution of $\text{LR}_M$.

In line with \citet{McLa:OnBo:1987}, \citet[][pp.~25--26]{McLa:Basf:mixt:1988}, \citet[][Sect.~6.6]{McLa:Peel:fini:2000} and \citet{Lo:Alik:2008}, $\text{LR}_M$ can be bootstrapped as follows.
Proceeding under $H_0^M$, a bootstrap sample is generated from model $M$ where $\boldsymbol{\vartheta}^M$ is replaced by its likelihood estimate formed under $H_0^M$ from the original sample.
The value of $\text{LR}_M$ is computed, for the bootstrap sample, after fitting models $M$ and VVV in turn to it.
This process is repeated independently $R$ times, and the replicated values of $\text{LR}_M$, formed from the successive bootstrap samples, provide an assessment of the null distribution of $\text{LR}_M$.
This distribution enables an approximation to be made to the $p$-value corresponding to the value of $\text{LR}_M$ evaluated from the original sample.

If a very accurate estimate of the $p$-value is required, then $R$ should be large \citep{Efro:Tibs:AnIn:1993}.
At the same time, when $R$ is large, the amount of computation involved is considerable.
However, there is usually no practical interest in estimating a $p$-value with high precision because the decision to be made concerns solely the rejection, or not, of $H_0^M$ at a specified significance level $\alpha$.

\citet{Aitk:Ande:Hind:Stat:1981} note that the bootstrap replications can be used to provide a test of approximate size $\alpha$.
In particular, the test that rejects $H_0^M$ if $\text{LR}_M$ for the original data is greater than the $h$th smallest of its $R$ bootstrap replications has size 
\begin{equation}
\alpha=1-\frac{h}{R+1},
\label{eq:significance level}
\end{equation}
approximately.
Hence, for a specified significance level $\alpha$, the values of $h$ and $R$ can be chosen according to \eqref{eq:significance level}.
For example, for $\alpha=0.05$, the smallest value of $R$ needed is 19 with $h=19$.
As cautioned above on the estimation of the $p$-value, $R$ needs to be large to ensure an accurate assessment.
In these terms, with $\alpha=0.05$, the value $R=99$ (and hence $h=95$) could be a good compromise \citep{McLa:OnBo:1987}.
  
\begin{figure}[!ht]
	\centering
		\resizebox{\textwidth}{!}{
		\begin{tabular}{c@{\hspace{-0.05mm}}c@{\hspace{-2mm}}c@{\hspace{-2mm}}c} 
		& $n=100$ & $n=200$ & $n=500$ \\[3mm]
		\begin{rotate}{90} \hspace{-0.4cm} $B=0.05$ \end{rotate}& 
		\begin{tabular}{c}
		\includegraphics[scale=0.33]{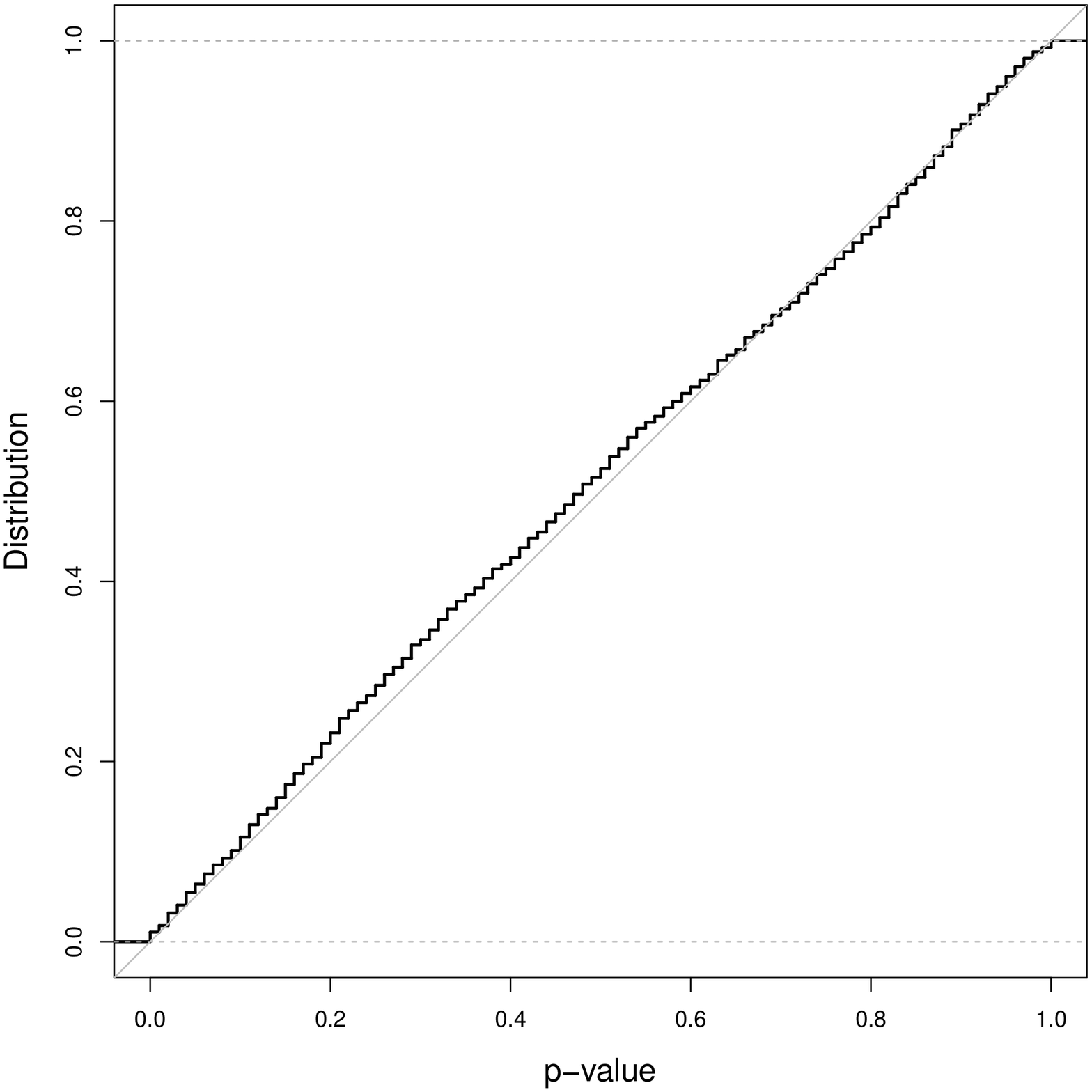} 
		\end{tabular}
		& 
		\begin{tabular}{c}
		\includegraphics[scale=0.33]{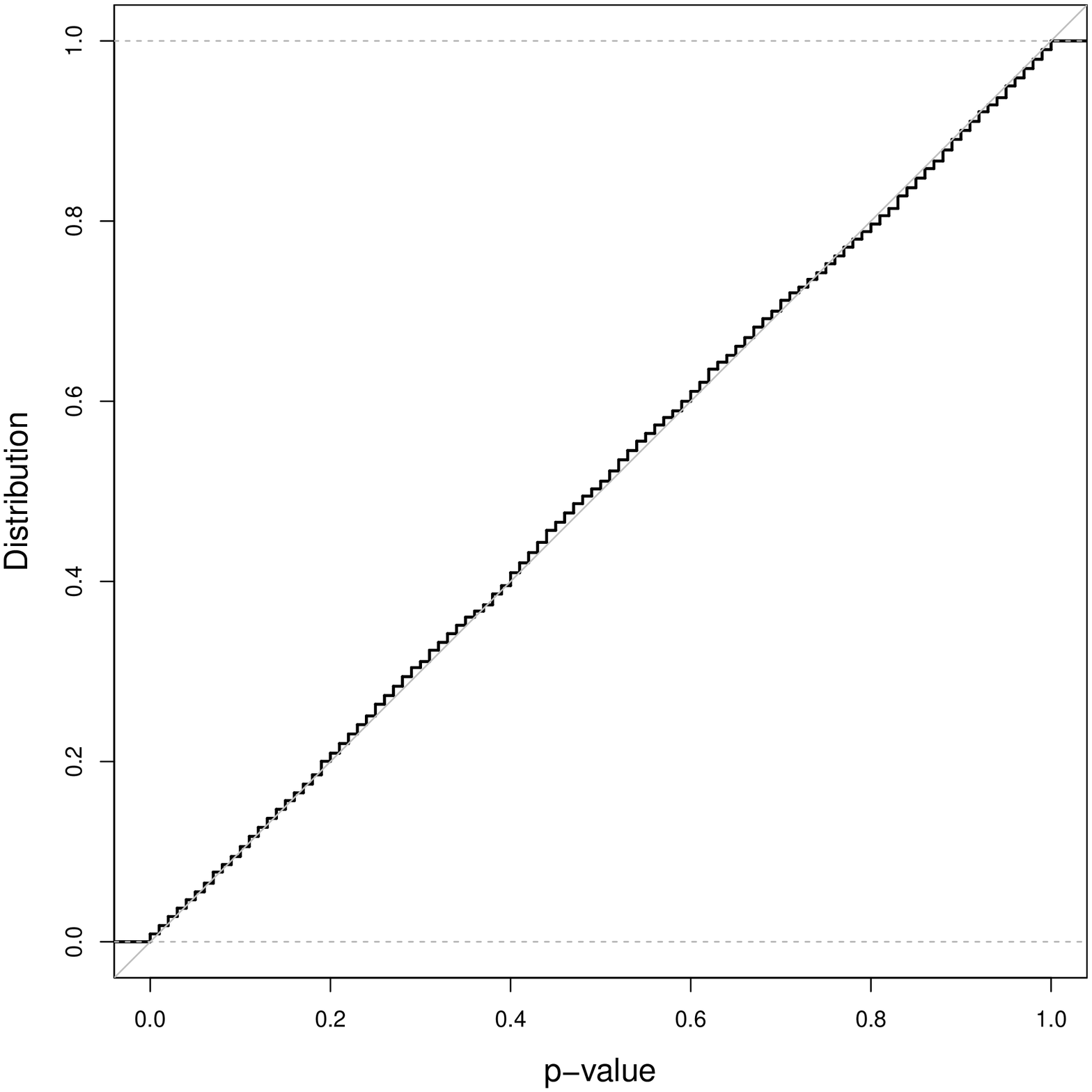} 
		\end{tabular}
		& 
		\begin{tabular}{c}
		\includegraphics[scale=0.33]{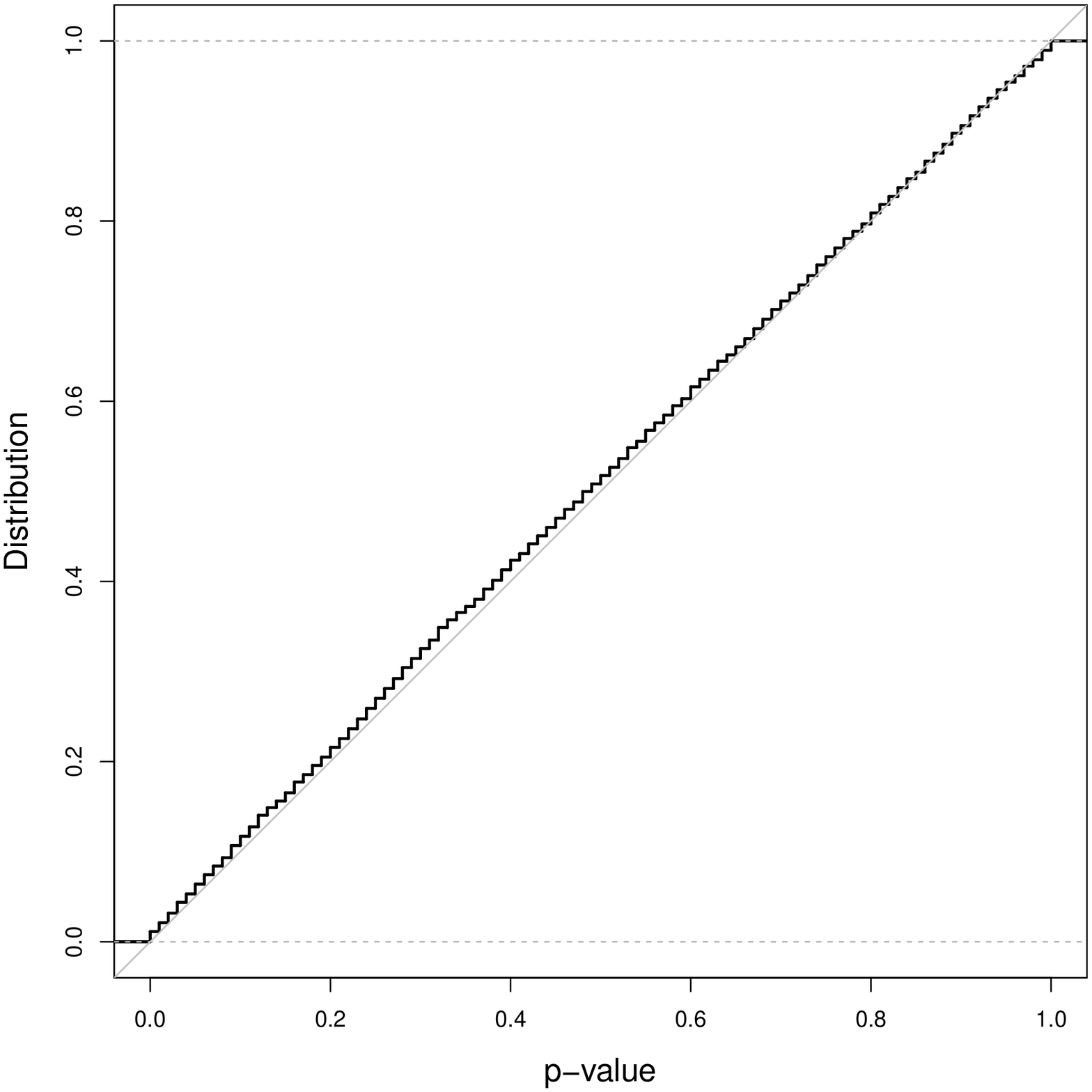} 
		\end{tabular}
		\\
		\begin{rotate}{90} \hspace{-0.39cm} $B=0.25$ \end{rotate}& 
		\begin{tabular}{c}
		\includegraphics[scale=0.33]{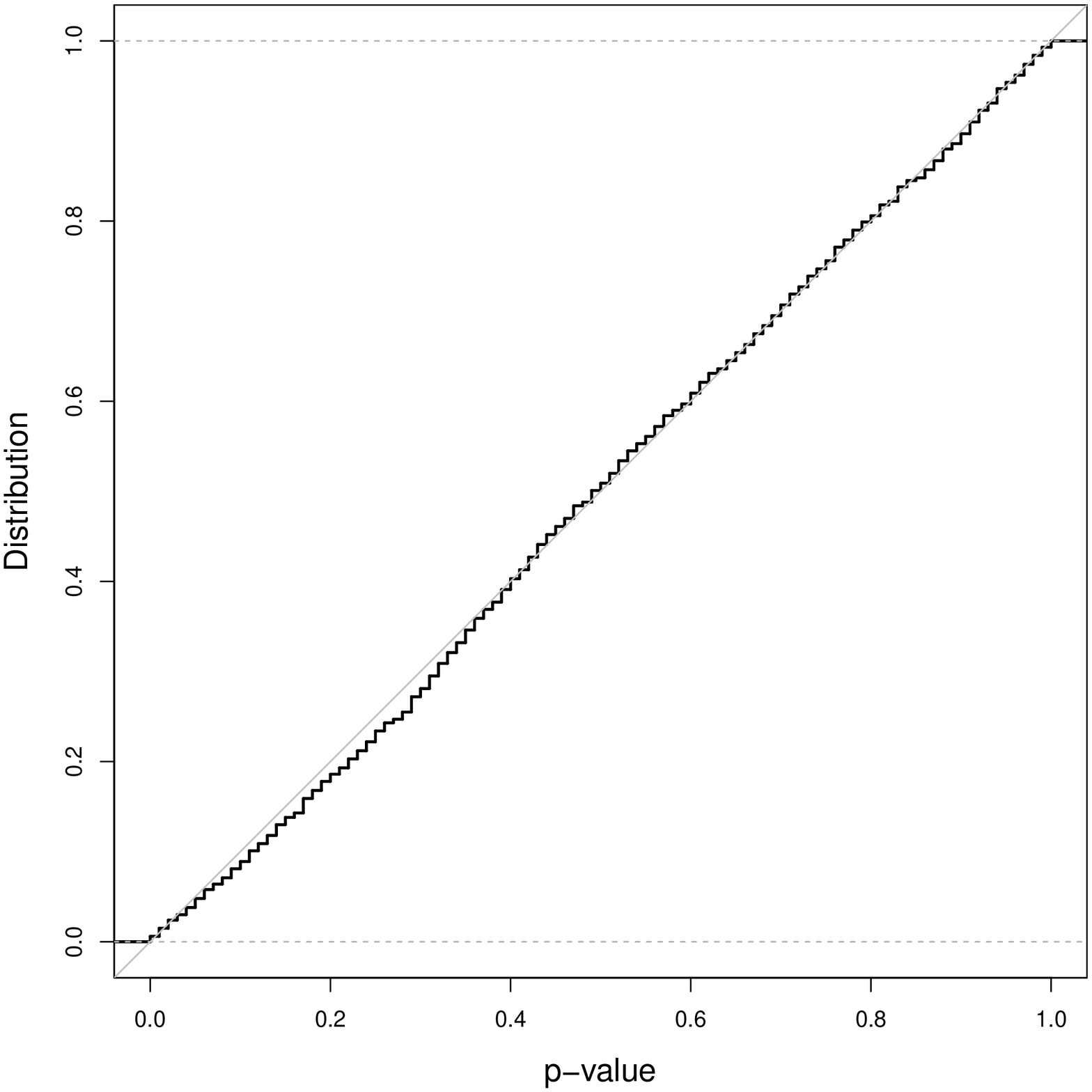} 
		\end{tabular}
		& 
		\begin{tabular}{c}
		\includegraphics[scale=0.33]{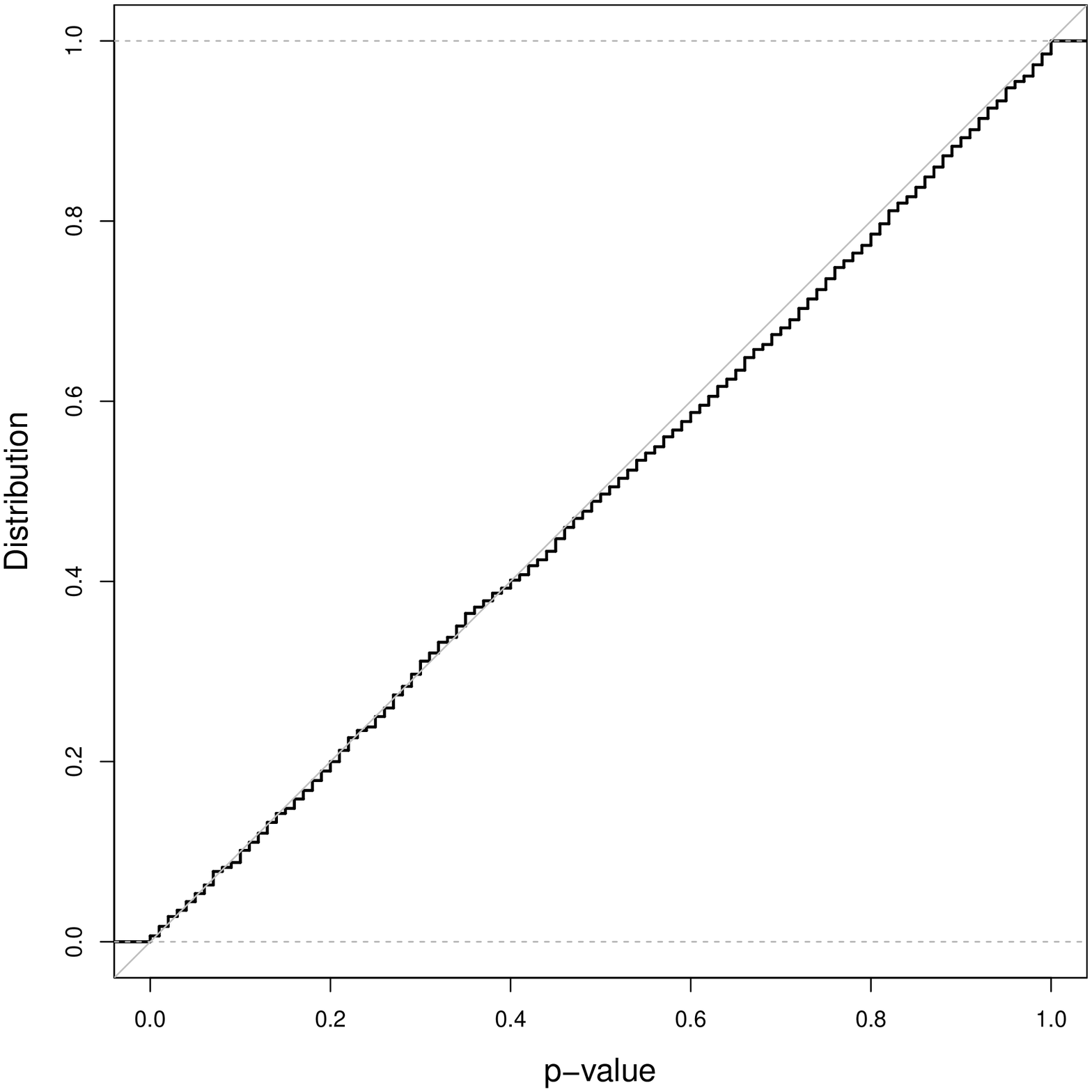} 
		\end{tabular}
		& 
		\begin{tabular}{c}
		\includegraphics[scale=0.33]{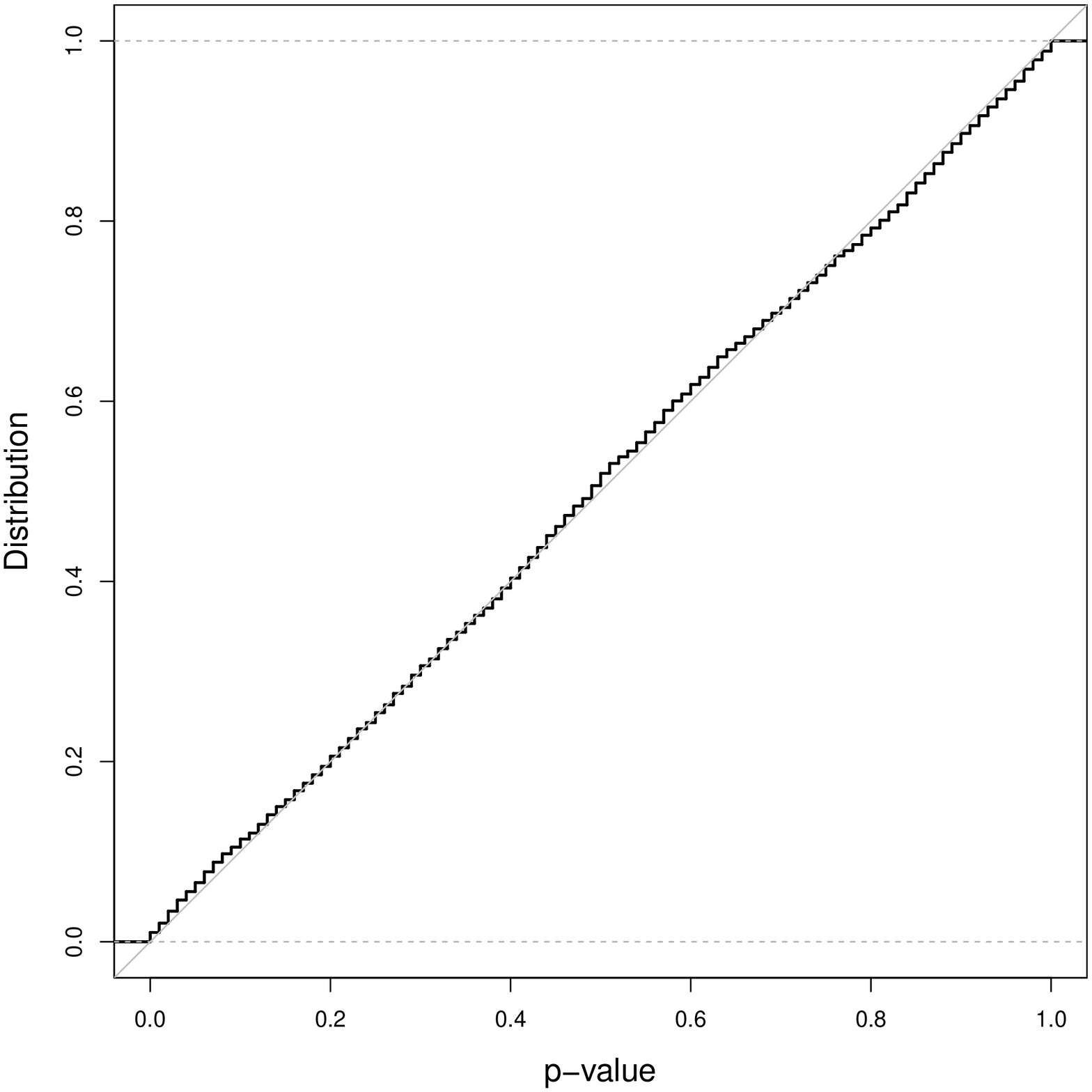} 
		\end{tabular}
		\\
		\begin{rotate}{90} \hspace{-0.40cm} $B=0.45$ \end{rotate}& 
		\begin{tabular}{c}
		\includegraphics[scale=0.33]{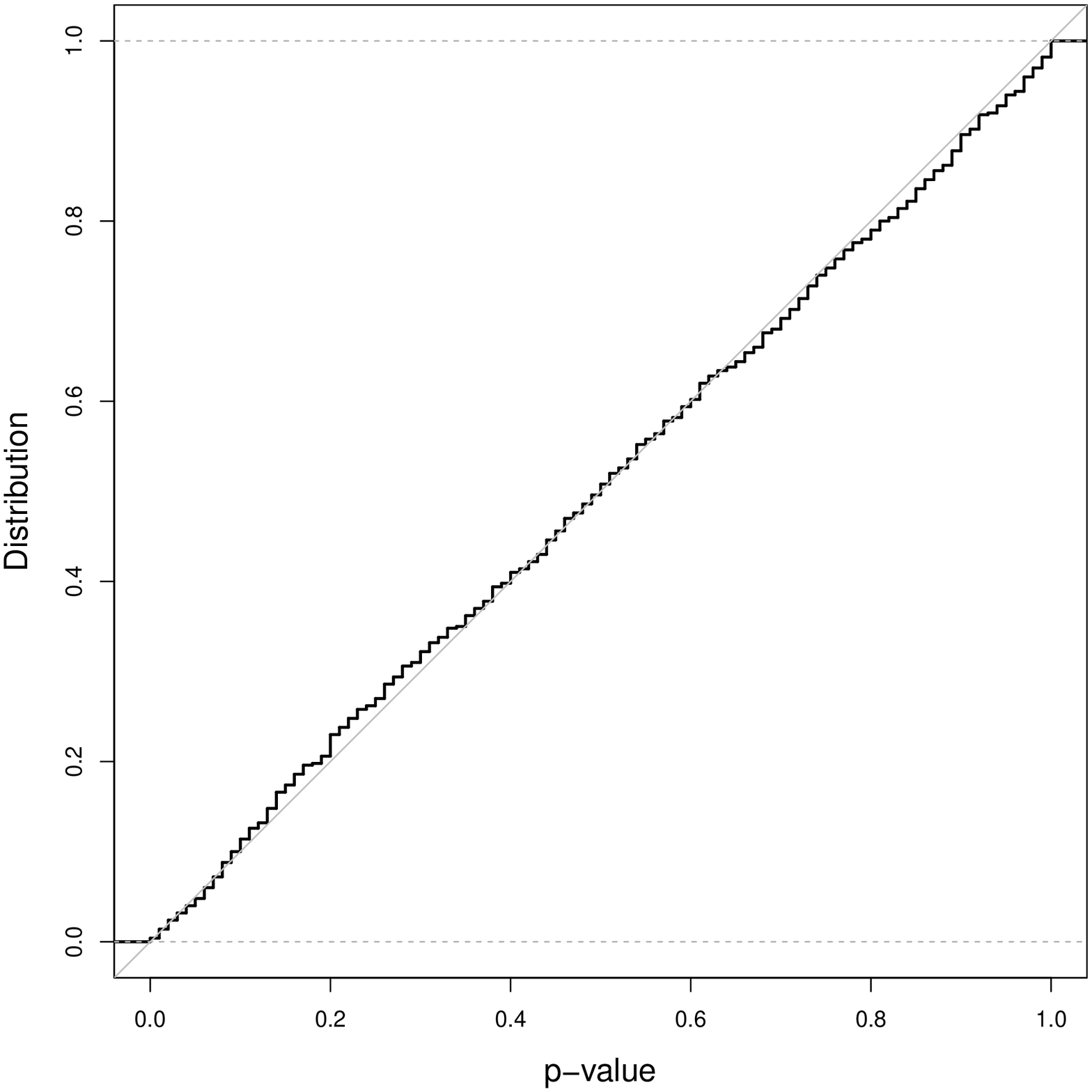} 
		\end{tabular}
		& 
		\begin{tabular}{c}
		\includegraphics[scale=0.33]{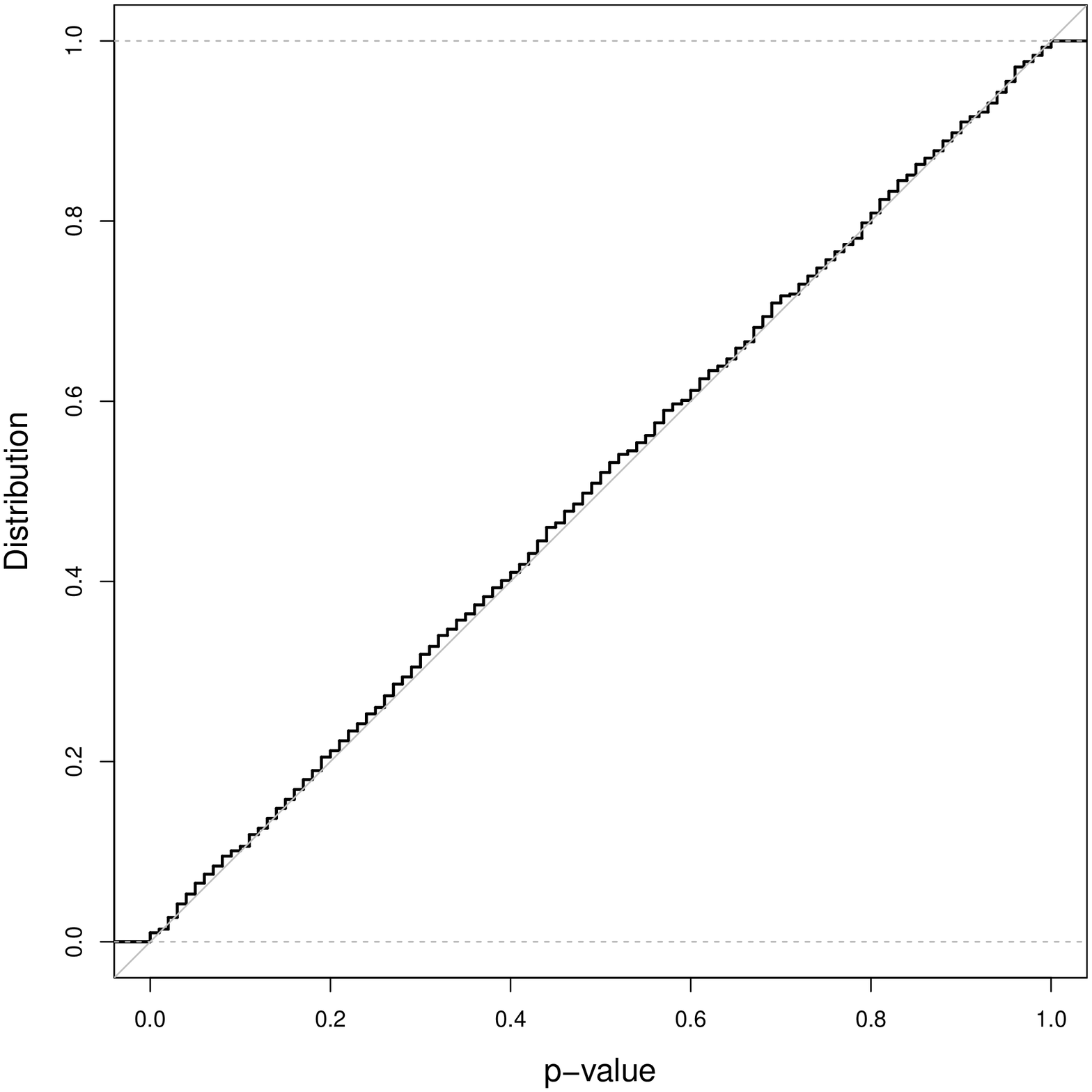} 
		\end{tabular}
		& 
		\begin{tabular}{c}
		\includegraphics[scale=0.33]{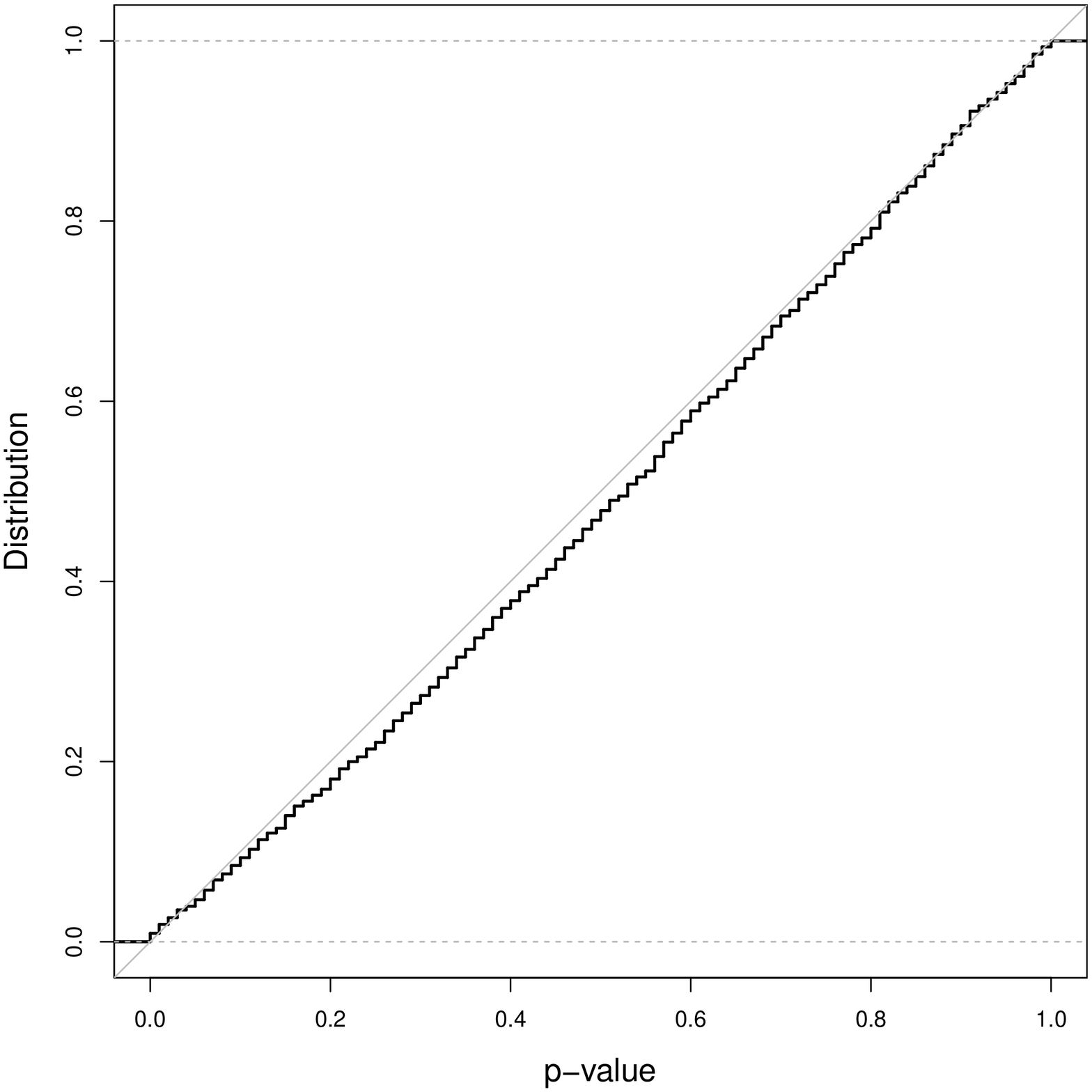} 
		\end{tabular}
	\end{tabular}
		}
	\caption{
	\footnotesize
	Parametric bootstrap: simulated distribution function of the $p$-values, for model EEE, at the varying of sample size $n$ and degree of overlap $B$.
	}
	\label{fig:EEE bootstrap}
\end{figure}

Under the same simulation design described in Sect.~\ref{subsec:Null distribution of the LR statistic}, \figurename~\ref{fig:EEE bootstrap} shows the results of the parametric bootstrap approach with $R=99$.
Furthermore, in this case, because the obtained results are very similar across models, we report only those referred to model EEE.
\figurename~\ref{fig:EEE bootstrap} illustrates thats the performance of the approach is always good regardless of the values of $n$ and $B$.

\section[Testing in the general family]{Testing in the general family}
\label{sec:Testing in the general family}

So far we have discussed the assessment of each model in $\overline{\mathcal{M}}$ with respect to the (benchmark) alternative model VVV, which is the most general, unconstrained model.
However, for real applications, we would prefer a statistical procedure to assess the true model with respect the entire general family $\widetilde{\mathcal{M}}$.
To this end we generalize, to the clustering context, the closed LR testing procedure proposed by \citet{Gres:Punz:Clos:2013} for completely labeled data.

\begin{figure}[!ht]
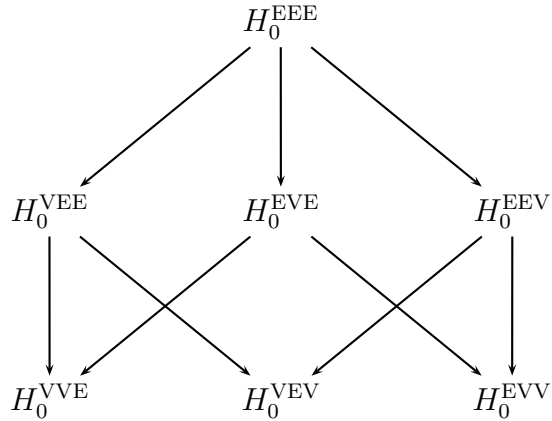

\centering
$
\begin{array}{c@{\hskip 2cm}c@{\hskip 2cm}c}
                          & \rnode{a}{H_0^\text{EEE}} &                           \\[2cm]
\rnode{b}{H_0^\text{VEE}} & \rnode{c}{H_0^\text{EVE}} & \rnode{d}{H_0^\text{EEV}} \\[2cm]
\rnode{e}{H_0^\text{VVE}} & \rnode{f}{H_0^\text{VEV}} & \rnode{g}{H_0^\text{EVV}} 
\end{array}
$
\psset{nodesep=3pt}
\everypsbox{\scriptstyle}
\ncline{->}{b}{e} 
\ncline{->}{c}{e} 
\ncline{->}{b}{f} 
\ncline{->}{d}{f} 
\ncline{->}{c}{g} 
\ncline{->}{d}{g} 
\ncline{->}{a}{b} 
\ncline{->}{a}{c} 
\ncline{->}{a}{d} 
\caption{Graph of the hierarchy of relationships between the null hypotheses.}
\label{fig:closed diagram}
\end{figure}

The hypotheses in 
$$
\mathcal{H}=\Bigl\{H_0^{\text{VVE}},H_0^{\text{VEV}},H_0^{\text{EVV}}\Bigr\}
$$ 
are said to be \textit{elementary} and, as detailed in \tablename~\ref{tab:elementary hypotheses}, they play a crucial role: depending on the true model in $\widetilde{\mathcal{M}}$, none, some, or all of the hypotheses in $\mathcal{H}$ may be the true null.
\begin{table}[!ht]
\caption{
\label{tab:elementary hypotheses}
Elementary hypotheses and their relationship with the models in $\widetilde{\mathcal{M}}$.}
	\centering
\begin{tabular}{c c c c c}
\toprule
$H_0^{\text{EVV}}$
& 
$H_0^{\text{VEV}}$
&
$H_0^{\text{VVE}}$
& 
&
True model
\\
\midrule
True & True & True  & $\Rightarrow$ & EEE \\[2mm]
False & True & True & $\Rightarrow$ & VEE \\[1mm]
True & False & True & $\Rightarrow$ & EVE \\[1mm]
True & True & False & $\Rightarrow$ & EEV \\[2mm]
False & False & True & $\Rightarrow$ & VVE \\[1mm]
False & True & False & $\Rightarrow$ & VEV \\[1mm]
True & False & False & $\Rightarrow$ & EVV \\[2mm]
False & False & False & $\Rightarrow$ & VVV \\
\bottomrule
\end{tabular}
\end{table}
\figurename~\ref{fig:closed diagram} also represents the null hypotheses $H_0^M$, $M\in\overline{\mathcal{M}}$, as a hierarchy where arrows indicate implications \citep[see][p.~344]{Hoch:Tamh:mult:1987}: for instance, $H_0^\text{EEE}$ implies $H_0^\text{VEE}$, and this also means that model EEE is more restrictive (i.e., more parsimonious) than model VEE.

Operationally, according to the closed LR testing procedure of \citet{Gres:Punz:Clos:2013}, we reject, say, the elementary hypothesis $H_0^{\text{VEV}}$ if and only if each LR test on the more restrictive hypotheses $H_0^{\text{VEE}}$, $H_0^{\text{EEV}}$, $H_0^{\text{EEE}}$, and also on $H_0^{\text{VEV}}$ itself, yields a significant result.
Denoting by $p_{\text{VEE}}$, $p_{\text{EEV}}$, $p_{\text{EEE}}$, and $p_{\text{VEV}}$ the $p$-values for $H_0^{\text{VEE}}$, $H_0^{\text{EEV}}$, $H_0^{\text{EEE}}$, and $H_0^{\text{VEV}}$, respectively, we report the \textit{adjusted} $p$-value for $H_0^{\text{VEV}}$ as $q_{\text{VEV}}=\max\left\{p_{\text{VEE}},p_{\text{EEV}},p_{\text{EEE}},p_{\text{VEV}}\right\}$.
An adjusted $p$-value represents the natural counterpart, in the multiple testing framework, of the classical $p$-value \citep[see, e.g.,][p.~18]{Bret:Hoth:West:Mult:2011}.
Specifically:
\begin{itemize}
	\item they provide information about whether $H_0\in\mathcal{H}$ is significant or not ($q_M$ can be compared directly with any chosen significance level $\alpha$ and if $q_M\leq\alpha$, then $H_0^M$ is rejected);
	\item they indicate ``how significant'' the result is (the smaller $q_M$, the stronger the evidence against $H_0^M$); and
	\item they are interpretable on the same scale as those for tests of individual hypotheses, making comparison with single hypothesis testing easier.
\end{itemize}

This closed testing procedure is the most powerful, among the available multiple testing procedures, that strongly controls the \textit{familywise error rate} (FWER) at level $\alpha$ \citep[as recently further corroborated via simulations by][]{GiAr:Bolz:Bonn:Cora:Solm:Adva:2012}.
Controlling the FWER in a strong sense means controling the probability of committing at least one Type I error under any partial configuration of true and false null hypotheses in $\mathcal{H}$. 
This is the only way to make inference on each hypothesis in $\mathcal{H}$.
For further details on the closed testing procedure, and on its properties, see \citet{Gres:Punz:Clos:2013}.

\section[Computational aspects]{Computational aspects}
\label{sec:Computational aspects}

Code for the LR tests (in both their $\chi^2$-based and bootstrap variants) and the closed LR testing procedure was written in the \textsf{R} computing environment \citep{R}. 
While specific code was written to obtain ML parameter estimation for models EVE and VVE (cf.\ Section~\ref{subsec:ML for EVE and VVE}), the \textbf{mixture} package \citep{Brow:McNi:mixt:2013} was used for the other models of the general family.

\subsection[Initialization]{Initialization}
\label{subsec:Initialization}

\subsubsection[LR tests]{LR tests}
\label{subsubsec:LR tests}

For model $M$, among the possible initialization strategies, each of the $n$ vectors $\boldsymbol{z}_{i}^{\left(0\right)}=(z_{i1}^{\left(0\right)},\ldots,z_{ik}^{\left(0\right)})'$ can be randomly generated either in a ``soft'' way by generating $k$ positive values summing to one, or in a ``hard'' way by randomly drawn a single observation from a multinomial distribution with probabilities $(1/k,\ldots,1/k)'$; see \citet{Bier:Cele:Gova:Choo:2003}, \citet{Karl:Xeka:Choo:2003}, and \citet{Bagn:Punz:Fine:2013} for more complicated strategies.
Let $\widehat{\boldsymbol{z}}_{i}^M$, $i=1,\ldots,n$, be the estimated posterior probabilities for model $M$.
Because model $M$ implies model VVV (that is $M$ is nested in VVV), the ``soft'' values of $\widehat{\boldsymbol{z}}_{i}^M$ can be used to initialize the EM algorithm for VVV; this forces, thanks to the monotonicity property of the EM algorithm \citep[see, e.g.,][]{McLa:Kris:TheE:2007}, $l_{\text{VVV}}$ to be greater than $l_M$ and, hence, $\text{LR}_M$ to be a well-defined positive value.   

In the generic bootstrap re-sample from the fitted model $M$ on the observed sample, we naturally know the true group membership of the generated observations.
Thus, we can use the corresponding true ``hard'' values of $\boldsymbol{z}_{i}$, $i=1,\ldots,n$, to initialize the EM algorithm for model $M$.
Once it is fitted, according to what said above, we can adopt the estimated posterior probabilities to initialize the EM algorithm for model VVV.      

\subsubsection[Closed testing procedure]{Closed testing procedure}
\label{subsubsec:Closed testing procedure}

With regard to the computation of the seven LR statistics in the closed testing procedure, on the observed sample we can take advantage of the hierarchy in \figurename~\ref{fig:closed diagram} to initialize the EM algorithms (for the use of hierarchical initialization strategies in mixture models see \citealp{Ingr:Mino:Punz:Mode:2013} and \citealp{Sube:Punz:Ingr:McNi:Clus:2013}).
In particular:
\begin{enumerate}
	\item a ``soft'' or ``hard'' random initialization is used for model EEE in the top of the hierarchy;
	\item the estimated posterior probabilities $\widehat{\boldsymbol{z}}_{i}^{\text{EEE}}$, $i=1,\ldots,n$, are used to initialize the EM algorithm for the models of the second level on the hierarchy (VEE, EVE, and EEV);
	\item the posterior probabilities of the model with the highest log-likelihood between VEE and EVE are used to initialize the EM algorithm for VVE; the posterior probabilities of the model with the highest log-likelihood between VEE and EEV are used to initialize the EM algorithm for VEV; the posterior probabilities of the model with the highest log-likelihood between EVE and EEV are used to initialize the EM algorithm for EVV;
	\item the posterior probabilities of the model with the highest log-likelihood between EVV, VEV, and VVE, are used to initialize the EM algorithm for VVV.
\end{enumerate}
The described hierarchical initialization guarantees the natural ranking of log-likelihoods $l_M$, $M\in\widetilde{\mathcal{M}}$.

\subsection{Convergence criterion}
\label{subsec:Convergence criterion}

The Aitken acceleration \citep{Aitk:OnBe:1926} is used to estimate the asymptotic maximum of the log-likelihood at each iteration of the EM algorithm. 
Based on this estimate, we can decide whether or not the algorithm has reached convergence, i.e., whether or not the log-likelihood is sufficiently close to its estimated asymptotic value. 
For model $M\in\widetilde{\mathcal{M}}$, the Aitken acceleration at iteration $q+1$, $q=0,1,\ldots$, is given by
\begin{equation*}
a_M^{\left(q+1\right)}=\frac{l_M^{\left(q+2\right)}-l_M^{\left(q+1\right)}}{l_M^{\left(q+1\right)}-l_M^{\left(q\right)}},
\end{equation*}
where $l_M^{\left(q+2\right)}$, $l_M^{\left(q+1\right)}$, and $l_M^{\left(q\right)}$ are the log-like\-li\-hood values from iterations $q+2$, $q+1$, and $q$, respectively. 
Then, the asymptotic estimate of the log-likelihood at iteration $q + 2$ is given by
\begin{equation*}	l_{M,\infty}^{\left(q+2\right)}=l_M^{\left(q+1\right)}+\frac{1}{1-a_M^{\left(q+1\right)}}\left(l_M^{\left(q+2\right)}-l_M^{\left(q+1\right)}\right),
\end{equation*}
cf.\ \citet{Bohn:Diet:Scha:Schl:Lind:TheD:1994}.
The EM algorithm can be considered to have converged when $l_{M,\infty}^{\left(q+2\right)}-l_M^{\left(q+1\right)}<\epsilon$ (see \citealp{Lind:Mixt:1995} and \citealp{McNi:Murp:McDa:Fros:Seri:2010}).

\section{Analysis on the Iris data}
\label{sec:Analysis on the Iris data}

In this section, we will show an application of the closed LR testing procedure on real data.
A nominal level of 0.05 is adopted for the FWER-control and $R=999$ bootstrap replications are considered; these values lead to $h=950$ in \eqref{eq:significance level}.
For completeness, the likelihood-based information criteria (IC) summarized in \tablename~\ref{tab:IC} will be also provided.
\begin{table}[!ht]
\caption{
\label{tab:IC}
Definitions and references for the adopted likelihood-based information criteria.
}
\centering
\begin{tabular}{*{3}c}
\toprule
IC
& 
Definition 
&  
Reference
\\
\midrule
AIC      & $2l_M-2\eta_M$                                                 & \citet{Akai:Info:1973}                         \\[3mm]
AIC$_3$  & $2l_M-3\eta_M$                                                 & \citet{Bozd:Theo:1994}
\\[3mm]
AICc     & $\text{AIC}-2\displaystyle\frac{\eta_M\left(\eta_M+1\right)}{n-\eta_M-1}$ & \citet{Hurv:Tsai:Regr:1989}          \\[3mm]
AICu     & $\text{AICc}-n\log\displaystyle\frac{n}{n-\eta_M-1}$             & \citet{McQu:Shum:Tsai:Them:1997}               \\[3mm]
AWE      & $2l_M-2\eta_M\left(\displaystyle\frac{3}{2}+\log n\right)$      & \citet{Banf:Raft:mode:1993}                    \\[3mm]
BIC      & $2l_M-\eta_M\log n$                                             & \citet{Schw:Esti:1978}                         \\[3mm]
CAIC     & $2l_M-\eta_M\left(1+\log n\right)$                              & \citet{Bozd:Mode:1987}                         \\[3mm]
ICL      & $\text{BIC} + \displaystyle\sum_{i=1}^n\sum_{j=1}^k \text{MAP}\left(\widehat{z}_{ij}^M\right)\log \widehat{z}_{ij}^M$ & 
\citet{Bier:Cele:Gova:Asse:2000}
\\
\bottomrule
\end{tabular}
\end{table}      
In the definition of the ICL, $\text{MAP}(\widehat{z}_{ij}^M)=1$, if $\max_{h=1,\ldots,k}\left\{\widehat{z}_{ih}^M\right\}$ occurs at component $j$, and $\text{MAP}(\widehat{z}_{ij}^M)=0$ otherwise.

\begin{figure}[!ht]
	\begin{center}		
\resizebox{0.9\textwidth}{!}{\includegraphics{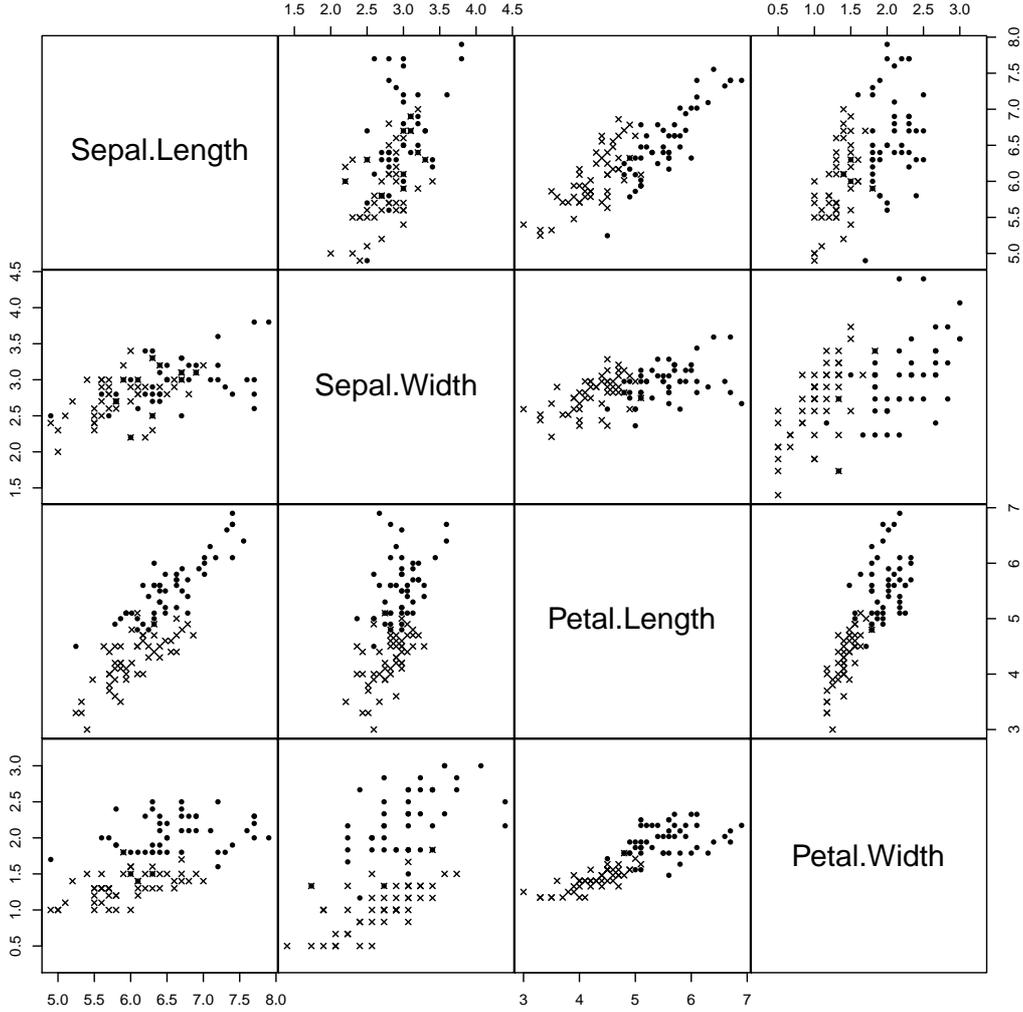}}
	\end{center}
	\vspace{-6mm}
	\caption{
	Matrix of scatter plots for a subset of Fisher's iris data	($\times$ denotes \textit{versicolor} and $\bullet$ denotes \textit{virginica})
	}
	\label{fig:iris}
\end{figure}

The Iris data set was made famous by \citet{Fish:Theu:1936} as an illustration of discriminant analysis.
Attention is here focused on the sample of $n=100$ iris subdivided in $k=2$ groups, of equal size, according to the species \textit{versicolor} and \textit{virginica}.
On these flowers, $p=4$ variables are measured in centimeters: \textit{sepal length}, \textit{sepal width}, \textit{petal length}, and \textit{petal width}.
The matrix of scatter plots for these data is shown in \figurename~\ref{fig:iris}.
A glance to this plot indicates that the groups are quite distinct, although some similarities in terms of orientation are observed.
This conjecture is represented by model VVE.

We now work in the case $k=2$ ignoring the known classification of the data.
\tablename~\ref{tab:IRIS - closed} contains the results from the closed LR testing procedure in both its $\chi^2$ and bootstrap versions.
\begin{table}[!ht]
\caption{
\label{tab:IRIS - closed}
Details on the closed LR testing procedure applied to the Iris data.
Bold numbers refer to the ``not rejected'' hypotheses in $\mathcal{H}$ at the 0.05-level (columns $q_M$).
}
\centering
\begin{tabular}{cccrcccccc}
\toprule
&&&&&\multicolumn{2}{c}{$\chi^2$-approximation}&&\multicolumn{2}{c}{bootstrap}\\
$M$ & \multicolumn{1}{c}{$\eta_M$} & \multicolumn{1}{c}{$\text{LR}_M$} & \multicolumn{1}{c}{$\nu_M$} && $p_M$ & $q_M$ && $p_M$ & $q_M$\\
\midrule
EEE & 19 & 39.38134 & 10 && 0.00002 &                  && 0.002 &                \\ 
VEE & 20 & 24.97787 &  9 && 0.00300 &                  && 0.012 &                \\ 
EVE & 22 & 25.40289 &  7 && 0.00064 &                  && 0.003 &                \\ 
EEV & 25 & 26.05177 &  4 && 0.00003 &                  && 0.001 &                \\ 
VVE & 23 & 10.70523 &  6 && 0.09793 & \textbf{0.09793} && 0.155 & \textbf{0.155} \\ 
VEV & 26 & 11.89078 &  3 && 0.00777 & 0.00777          && 0.026 & 0.026          \\ 
EVV & 28 & 10.93548 &  1 && 0.00094 & 0.00094          && 0.010 & 0.010          \\ 
VVV & 29 &          &    &&         &                  &&       &                \\ 
\bottomrule
\end{tabular}
\end{table}
First of all, we note that the null hypothesis $H_0^{\text{EEE}}$ is rejected at any reasonable level ($p_{\text{EEE}}=0.00002$ with the $\chi^2$-approximation, and $p_{\text{EEE}}=0.002$ with the bootstrap approximation).
Hence, if we limit the attention to this test only, as it is typically done in the literature, we should lean towards the adoption of a heteroscedastic Gaussian mixture characterized by 29 parameters.      
On the contrary, additional information can be gained by looking at the closed testing procedure.
In particular, \tablename~\ref{tab:IRIS - closed} lists unadjusted $p$-values, for all the hypotheses in the hierarchy, and adjusted $p$-values for the elementary hypotheses.
To facilitate the comprehension of how the adjusted $p$-values are computed, we can consider the following example in the bootstrap case: the adjusted $p$-value for $H_0^{\text{VEV}}$ is given by
\begin{eqnarray*}
q_{\text{VEV}}&=&\max\bigl\{p_{\text{VEV}},p_{\text{EEV}},p_{\text{VEE}},p_{\text{EEE}}\bigr\}\\
&=&\max\bigl\{0.026,0.001,0.012,0.002\bigr\}=0.026.
\end{eqnarray*}

At the 0.05-level, because $H_0^\text{VVE}$ is the only elementary hypothesis that is not rejected in $\mathcal{H}$, it is also the hypothesis to be retained at the end of the procedure (see \tablename~\ref{tab:elementary hypotheses}).
This result does not vary by varying the approximation ($\chi^2$ or bootstrap) of the LR statistics and it confirms our graphical conjectures.
Moreover, it allows us to obtain a more parsimonious model (having 23 parameters) with a gain of 6 parameters with respect to model VVV.
Note also that, both the models (VVE and VVV) lead to five misallocated observations.

\tablename~\ref{tab:IRIS - IC} reports the values of the likelihood-based information criteria of \tablename~\ref{tab:IC} for these data.
\begin{table}[!ht]
\caption{
\label{tab:IRIS - IC}
Likelihood-based information criteria for the Iris data.
Bold numbers refer to the best model (highest column value) for each information criterion.
}
\centering
\resizebox{\textwidth}{!}{
\begin{tabular}{ccccccccccc}
\toprule
$M$ & \multicolumn{1}{c}{$\eta_M$} & \multicolumn{1}{c}{$2l_M$} & \multicolumn{1}{c}{AIC} & \multicolumn{1}{c}{AIC$_3$} & \multicolumn{1}{c}{AICc} & \multicolumn{1}{c}{AICu} & \multicolumn{1}{c}{AWE} & \multicolumn{1}{c}{BIC} & \multicolumn{1}{c}{CAIC} & \multicolumn{1}{c}{ICL} \\
\midrule
EEE & 19 & -298.63 & -336.63 & -355.63 & -346.13 & -368.45 & -530.63 & -386.13 & -405.13 & -390.18 \\ 
VEE & 20 & -284.23 & -324.23 & -344.23 & -334.86 & -358.43 & \textbf{-528.43} & -376.33 & \textbf{-396.33} & -378.62 \\ 
EVE & 22 & -284.65 & -328.65 & -350.65 & -341.80 & -367.93 & -553.28 & -385.97 & -407.97 & -390.54 \\ 
EEV & 25 & -285.30 & -335.30 & -360.30 & -352.87 & -382.98 & -590.56 & -400.43 & -425.43 & -404.13 \\ 
VVE & 23 & -269.96 & \textbf{-315.96} & \textbf{-338.96} & \textbf{-330.48} & \textbf{-357.93} & -550.79 & \textbf{-375.87} & -398.87 & \textbf{-378.24} \\ 
VEV & 26 & -271.14 & -323.14 & -349.14 & -342.37 & -373.84 & -588.61 & -390.88 & -416.88 & -392.80 \\ 
EVV & 28 & -270.19 & -326.19 & -354.19 & -349.06 & -383.31 & -612.07 & -399.13 & -427.13 & -402.90 \\ 
VVV & 29 & -259.25 & -317.25 & -346.25 & -342.11 & -377.77 & -613.35 & -392.80 & -421.80 & -394.40 \\ 
\bottomrule
\end{tabular}
}
\end{table}
Some concern arises when noting how different criteria can lead to different choices; this consideration is further exacerbated if we consider that practitioners tend to use one of them almost randomly or routinely. 
On the other hand, the closed LR testing procedure offers a straightforward assessment of the model in the general family and it is based on only one subjective element, the significance level $\alpha$, whose meaning is clear to everyone.
Moreover, the adjusted $p$-values also provide a measure of ``how significant'' the test result is for each of the three terms of the eigen-decomposition: volume, shape, and orientation.

\section{Discussion and future work}
\label{sec:Concluding remarks and discussion}

The likelihood-ratio statistic for comparing the homoscedastic Gaussian mixture \textit{versus} its heteroscedastic version has been studied only in the univariate case \citep[see][]{Lo:Alik:2008}.
Even if it were generalized to the multivariate case, being the resulting test \textit{omnibus}, the practitioner should remain without any further information about a possible similarity across groups, different from the homoscedastic one, if the corresponding null hypothesis were rejected.
Motivated by these considerations, in this paper we extended the use of likelihood-ratio tests in the multivariate case and to the general family of eight Gaussian mixture models of \citet{Cele:Gova:Gaus:1995}, homoscedasticity and heteroscedasticity being the extreme configurations in parsimony terms.

For two of the models in the general family, we also derived maximum likelihood parameter estimates fulfilling the requirements of the family --- this is above and beyond the work of \citet{Cele:Gova:Gaus:1995}.
For the resulting seven tests, we presented simulation results which were in line with those obtained by \citet{Lo:Alik:2008} in the univariate case for the likelihood-ratio test of homoscedasticity for Gaussian mixtures: the $\chi^2$ reference distribution under the null does not provide a reasonable approximation for the likelihood-ratio statistic for small sample sizes and/or for large overlap between groups.
To overcome this problem, we adopted a parametric bootstrap approach.
Following work of \citet{Gres:Punz:Clos:2013} in the completely labeled scenario, the obtained tests were also simultaneously considered in a closed testing procedure in order to assess a choice in the whole general family.

Although, in principle, an information criterion could be employed, a large number of these criteria have been proposed in literature (possibly leading to different choices as shown in the application to real data) and practitioners tend to use a given one of them routinely.
On the other hand, the closed testing procedure illustrated here offers a straightforward assessment of the model in the general family and it is only based on one subjective element, the significance level $\alpha$, whose meaning is clear to everyone.
The real data set analyzed in the paper showed the gain in information and in parsimony that can be obtained by this approach.

A further remark refers to the type of application that is not restricted to model-based clustering.
Our proposal provides indeed a suitable way to assess the model in the general family also for model-based classification --- naturally based on Gaussian mixtures --- where we fit our mixture models to data where some of the observations have known labels.
In this case, we have also the advantage to know in advance the number of groups.
Future work will investigate the extension of the closed testing procedure to the analogue general family for mixtures of $t$ distributions \citep{Andr:McNi:Mode:2012} and for mixtures of contaminated Gaussian distributions \citep{Punz:McNi:Outl:2013}. 

\bibliographystyle{natbib}
\bibliography{References}

\end{document}